\documentclass[useAMS,usenatbib]{mn2e}
\usepackage{graphicx}
\usepackage{lscape}
\title[Ultra-cool dwarfs in wide binary systems]{Discovery of the first wide L dwarf + giant binary system and eight other ultra-cool dwarfs in wide binaries}
\author[Z.H. Zhang et al.]{Z.H. Zhang$^{1}$\thanks{E-mail:
zenghuazhang@hotmail.com},
D. J. Pinfield$^{1}$, A. C. Day-Jones$^{2,1}$, B. Burningham$^{1}$,
\newauthor
 H. R. A. Jones$^{1}$, S. Yu$^{3}$, J. S. Jenkins$^{2}$, Z. Han$^{4,7}$,  M. C. G\'{a}lvez-Ortiz$^{1}$, 
\newauthor 
J. Gallardo$^{2}$, A. E. Garc\'{i}a P\'{e}rez$^{5,1}$,  D. Weights$^{1}$, C. G. Tinney$^{6}$, R. S. Pokorny$^{4,7}$ \\
$^{1}$Centre for Astrophysics Research, Science and Technology Research Institute, University of Hertfordshrie, Hatfield AL10 9AB, UK\\
$^{2}$Department of Astronomy, Universidad de Chile, Casilla postal 36D, Santiago, Chile\\
$^{3}$Armagh Observatory, College Hill, Armagh BT61 9DG, Northern Ireland, UK \\
$^{4}$National Astronomical Observatories/Yunnan Observatory, Chinese Academy of Sciences, Kunming 650011, China\\
$^{5}$Department of Astronomy, University of Virginia, P.O. Box 400325, Charlottesville, VA 22904-4325, USA \\
$^{6}$School of Physics, University of New South Wales, 2050, Australia \\
$^{7}$Key Laboratory for the Structure and Evolution of Celestial Objects, Chinese Academy of Sciences, China\\
}
\begin{document}

\date{Accepted 2009 ------ ---. Received 2009 October 4; in original form 2009 June 16}

\pagerange{\pageref{firstpage}--\pageref{lastpage}} \pubyear{2002}

\maketitle

\label{firstpage}

\begin{abstract}
We identify 806 ultra-cool dwarfs from their SDSS \emph{riz} photometry 
(of which 34 are newly discovered L dwarfs) and obtain proper motions through 
cross matching with UKIDSS and 2MASS. Proper motion and distance constraints show 
that nine of our ultra-cool dwarfs are members of widely separated binary systems; 
SDSS 0101 (K5V+M9.5V), SDSS 0207 (M1.5V+L3V), SDSS 0832 (K3III+L3.5V), SDSS 0858 
(M4V+L0V), SDSS 0953 (M4V+M9.5V), SDSS 0956 (M2V+M9V), SDSS 1304 (M4.5V+L0V), 
SDSS 1631 (M5.5V+M8V), SDSS 1638 (M4V+L0V). One of these (SDSS 0832) is shown to 
be a companion to the bright K3 giant $\eta$ Cancri. Such primaries can provide 
age and metallicity constraints for any companion objects, yielding excellent 
benchmark objects. $\eta$ Cancri AB is the first wide ultra-cool dwarf + giant 
binary system identified. We present new observations and analysis that constrain 
the metallicity of $\eta$ Cancri A to be near solar, and use recent evolutionary 
models to constrain the age of the giant to be 2.2$-$6.1 Gyr. If $\eta$ Cancri B 
is a single object, we estimate its physical attributes to be; mass = 63$-$82
M$_{Jup}$, $T_{\rm eff}$ = 1800$\pm$150 K, log $g$ = 5.3$-$5.5, [M/H] = 0.0$\pm$0.1. 
Its colours are non typical when compared to other ultra-cool dwarfs, and we also 
assess the possibility that $\eta$ Cancri B is itself an unresolved binary, showing 
that the combined light of an L4 + T4 system could provide a reasonable explanation 
for its colours.

\end{abstract}


\section{Introduction}
The number of discovered brown dwarfs is increasing rapidly (e.g., DwarfArchives.org), benefitting from optical and near-infrared large area surveys such as the Deep Near-Infrared Survey of the Southern Sky \citep[DENIS;][]{epc}, the Two Micron All Sky Survey \citep[2MASS;][]{skr}, the Sloan Digital Sky Survey \citep[SDSS;][]{aba}, and the UKIRT (United Kingdom Infrared Telescope) Infrared Deep Sky Survey \citep[UKIDSS;][]{law}. Ultra-cool dwarfs have red optical-near-infrared colours due to their low effective temperature. Observable ultra-cool dwarfs are nearby because of their intrinsic faintness, and can therefore have relatively higher proper motion. It is very effective to select ultra-cool dwarfs by colour \citep[e.g.][]{cru,chi,zh09b,pinf} or proper motion \citep{dea,she}. Proper motion is also a powerful tool for identifying ultra-cool dwarf companions to other stars since such binary systems are common proper motion pairs \citep[e.g.][]{bur,day,luh,pin,tok}. We generically refer to such ultra-cool dwarfs as benchmark objects, for which optimal primaries include subgiant/giant stars and white dwarfs \citep{pin} along with the more common main sequence primaries \citep{jen}. Such systems are valuable, since ultra-cool dwarf properties may be constrained (at some level) by the primary star.

Current brown dwarf models have difficulty in accurately reproducing
observations, and benchmark objects are thus needed to calibrate both
atmospheric and evolutionary models. In order for a brown dwarf to be
considered a benchmark it must have one or more properties (e.g. age,
mass, distance, metallicity) that can be constrained relatively independently of ultra-cool dwarf
models. The overall usefulness of an object as a benchmark is also
dependent on the accuracy of the measured properties and the number
of assumptions that have to be made if some degree of referencing to
models is required.

The galactic disk population provides a number of host environments
that are useful for the discovery of such benchmarks. Many identified benchmarks are members of young clusters such as the Hyades, Pleiades and Praesepe \citep{bou,hog,case,bih,reb, cha, mag,mar98,mar96,zap,cos} and moving groups \citep[e.g. LP 944-20;][]{rib,tin,cla}. These kinds of benchmarks provide accurate ages and metallicities that can be inferred from other cluster/group members. However due to cluster evaporation \citep{bou} or disk heating mechanisms \citep{sim}, the spatial concentrations become dispersed and kinematic signatures diffused after $\sim$1 Gyr. Such benchmarks thus only populate the $<$1 Gyr age range, and are not fully applicable for studies of brown dwarf evolution across the full age extent of the galactic disk. This is also the case for isolated field brown dwarfs that have Lithium in their atmospheres, such as DENISp-J1228.2-1547 \citep{del97,tin97}, SDSSJ0423-0414 \citep{bur05}, 2M0850 \citep{kir99} and Kelu1 \citep{rui97}, where the Lithium test can be used to estimate their age \citep{mag93}.

More useful for this purpose are brown dwarfs as members of binaries, where
the host star can provide age, distance and in some cases
metallicity constraints. How useful these systems are is dependent on how
well we understand the nature and physics of the primary stars.
Main sequence star ages can be largely
uncertain due to the degeneracy of evolutionary models on the
main-sequence \citep[e.g.][]{gir,yi}. Tighter age constraints can be gained from more evolved
stars, e.g. subgiants, giants and white dwarfs. Such primaries can constrain accurate ages, as well
as metallicity in the case of subgiants and early phase giant stars, via
robust models. These types of binary systems can populate the full age range of the disk up 
to 10 Gyrs \citep[see][]{pin}.

Benchmark brown dwarfs can also be in brown dwarf + brown dwarf binary
pairs where the dynamical mass of the components can be
calculated. \citet{liu8} suggest that these systems will
yield good determinations of gravity and
age. However they cannot, in general provide
accurate metallicity estimations, except for the rare examples that are found
to be members of a cluster or components of a higher multiple system (where
a higher mass component has known metallicity), as is the case for HD 10948BC \citep{dup}.

In this paper we report the discovery of nine common proper motion binary systems with ultra-cool dwarf components. One is a benchmark L dwarf companion to the giant star $\eta$ Cancri, and the other eight are late M and  early L dwarf companions to K-M dwarf stars. The photometric selection processes are presented in Section 2. The spectral types are presented in Section 3. Distances are presented in Section 4. Proper motions are presented in Section 5. $\eta$ Cancri AB is presented and discussed in Section 6, and Section 7 presents a summary.

\section{Photometric selection}

\subsection{SDSS identification}
The SDSS uses a dedicated 2.5 m telescope
equipped with a large format mosaic CCD camera to image the sky in
five optical bands (\emph{u}, \emph{g}, \emph{r}, \emph{i},
\emph{z}), and two digital spectrographs to obtain the spectra of
galaxies, quasars and late type stars selected from the imaging data
\citep{yor}. The SDSS magnitude limits (95\%
detection repeatability for point sources) for the \emph{u},
\emph{g}, \emph{r}, \emph{i} and \emph{z} bands are 22.0, 22.2,
22.2, 21.3 and 20.5 respectively, all magnitudes are in SDSS AB system \citep{ade}. 

\begin{figure} %  figure placement: here, top, bottom, or page
   \centering
   \includegraphics[width= \columnwidth]{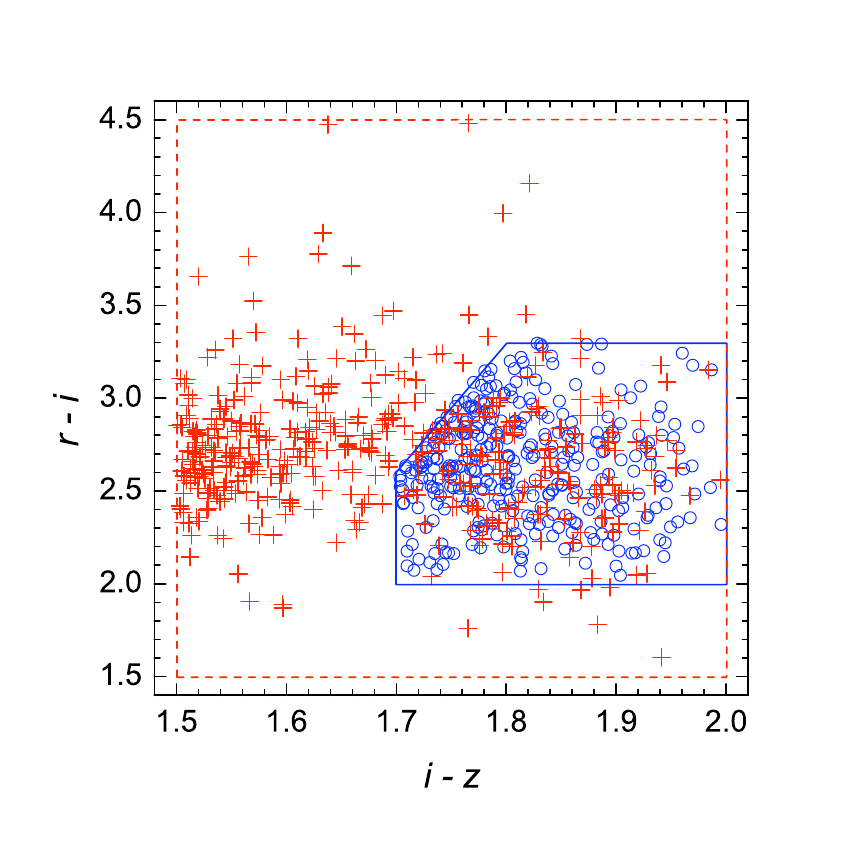}
   \caption{\emph{riz} colour space for ultra-cool dwarfs with (\emph{red crosses}) and without (\emph{blue circles}) SDSS spectra. The \emph{polygons} show the \emph{r-i} and \emph{i-z} colour limits used to select ultra-cool dwarfs with (\emph{red dashed lines}) and without (\emph{blue solid lines}) spectra.}
   \label{riz}
\end{figure}

The SDSS $i-z$ colour is particularly useful for L dwarf selection \citep{fan}.
We used $2<i-z<3.2$ in our previous ultra-cool proper motion work  \citep{zh09b} to select L dwarf candidates. In this work, we focus on a bluer sample to increase our sample size and increase the fraction of objects that have measured SDSS spectra. We used  $1.5<i-z<2$ and $1.5<r-i<4.5$ for objects with SDSS spectra, leading to a total ultra-cool dwarf sample of 3154 objects. For objects without spectra we used a more constraining selection, requiring $1.7<i-z<2, 2< r-i<3.3, r-i<7(i-z)-9.3, 16<z<19.5$, and $17<i<21.2$. Figure \ref{riz} shows the colour criteria for these selections. 

We also made a deeper ultra-cool dwarf search towards the young open cluster Praesepe \citep[M44; age = 0.79 Gyr, d = 181.5 pc;][]{va2}. The original motivation was to identify faint cluster members, however the most interesting discovery from this search was a foreground field object (see Section 5.2), so we have chosen to present the analysis of this deeper search here, along with that of our shallower search in the field. The Praesepe cluster has been surveyed by SDSS and UKIDSS (the Galactic Cluster Survey, within 3$^{\circ}$ of 08h40m06s,+19$^{\circ}$41$\arcmin$06$\arcsec$), and we used the same colour criteria as previously described for SDSS objects without spectra. However, the original aim to identify ultra-cool dwarfs at the distance of the cluster \citep[e.g.][]{va2,tay06} led us to limit this search to $19<z<20.5$ and $21<i<22.4$, corresponding to early L dwarfs \citep[$i-z$ = 1.8$-$2.1;][]{zh09b} at $\sim$100$-$200 pc (see the M$_{i} - (i-z)$ spectral type relationship \citep{haw}).

\subsection{Near-infrared photometry}
Near-infrared colours provide additional information useful for spectral typing ultra-cool dwarfs. SDSS selected candidates were cross matched with point sources in the 2MASS \citep{cut} and UKIDSS \citep{law} catalogues. We used a matching radius of 6$^{\prime\prime}$ to ensure that ultra-cool dwarfs with high proper motion would generally be matched by this procedure, despite possible motion over a period of up to $\sim$8 years (between SDSS-2MASS, and SDSS-UKIDSS epochs).  563 objects with spectra were cross matched in SDSS and 2MASS, of which 469 were also cross matched in UKIDSS. 1761 objects without spectra were cross matched in SDSS and 2MASS, of which 337 objects were cross matched in UKIDSS. So 806 SDSS objects (see online material Table 9 for these ultra-cool dwarfs) were cross matched in 2MASS and UKIDSS of which 469 objects have SDSS spectra. For our ultra-cool dwarf sample towards Praesepe, we cross matched candidates with the UKIDSS Galactic Cluster Survey (GCS), which covers 28 square degrees of the cluster (radius $\sim$3$^{\circ}$) in \emph{ZYJHK}.

\begin{figure*} %  figure placement: here, top, bottom, or page
   \centering
   \includegraphics[width=530pt]{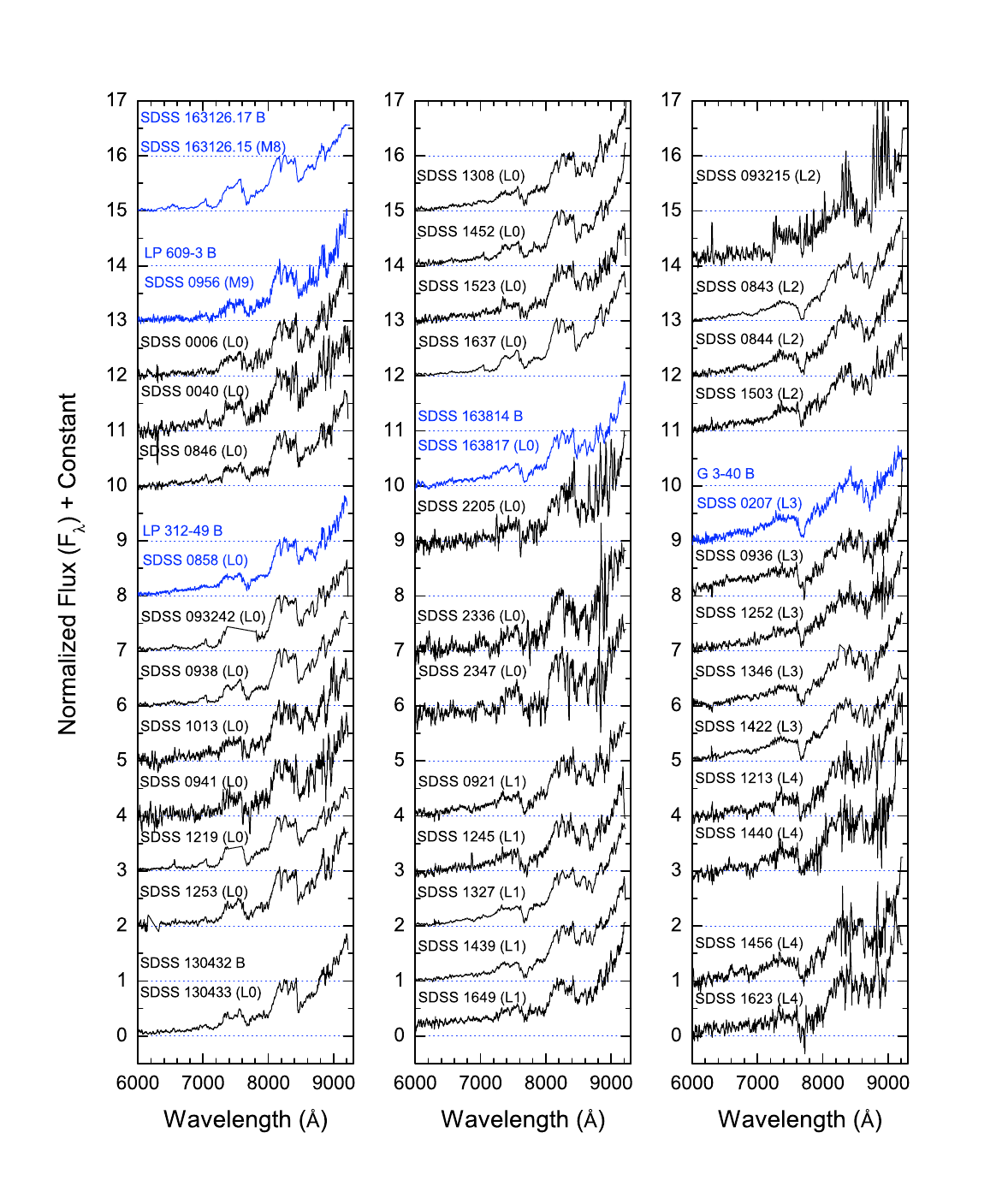}
   \caption{SDSS spectra of 39 ultra-cool dwarfs. Six objects in binary systems are those with two names (see Section \ref{s5.3}), of which five are previously identified ultra-cool dwarfs (\emph{blue}). The remaining 34 objects (\emph{black}) are new L dwarfs discovered in this work, one of which is in a binary system. Note that two of the binaries we present in this paper do not have spectroscopy at this time. All spectra have been normalized to one at 8250 {\AA}, smoothed by 11 pixels, and vertically offset for clarity. \emph{Dotted lines} indicate Zero point offsets and also normalization levels to aid the visual examination of spectra.}
   \label{s39}
\end{figure*}

\begin{figure} %  figure placement: here, top, bottom, or page
   \centering
   \includegraphics[width= \columnwidth]{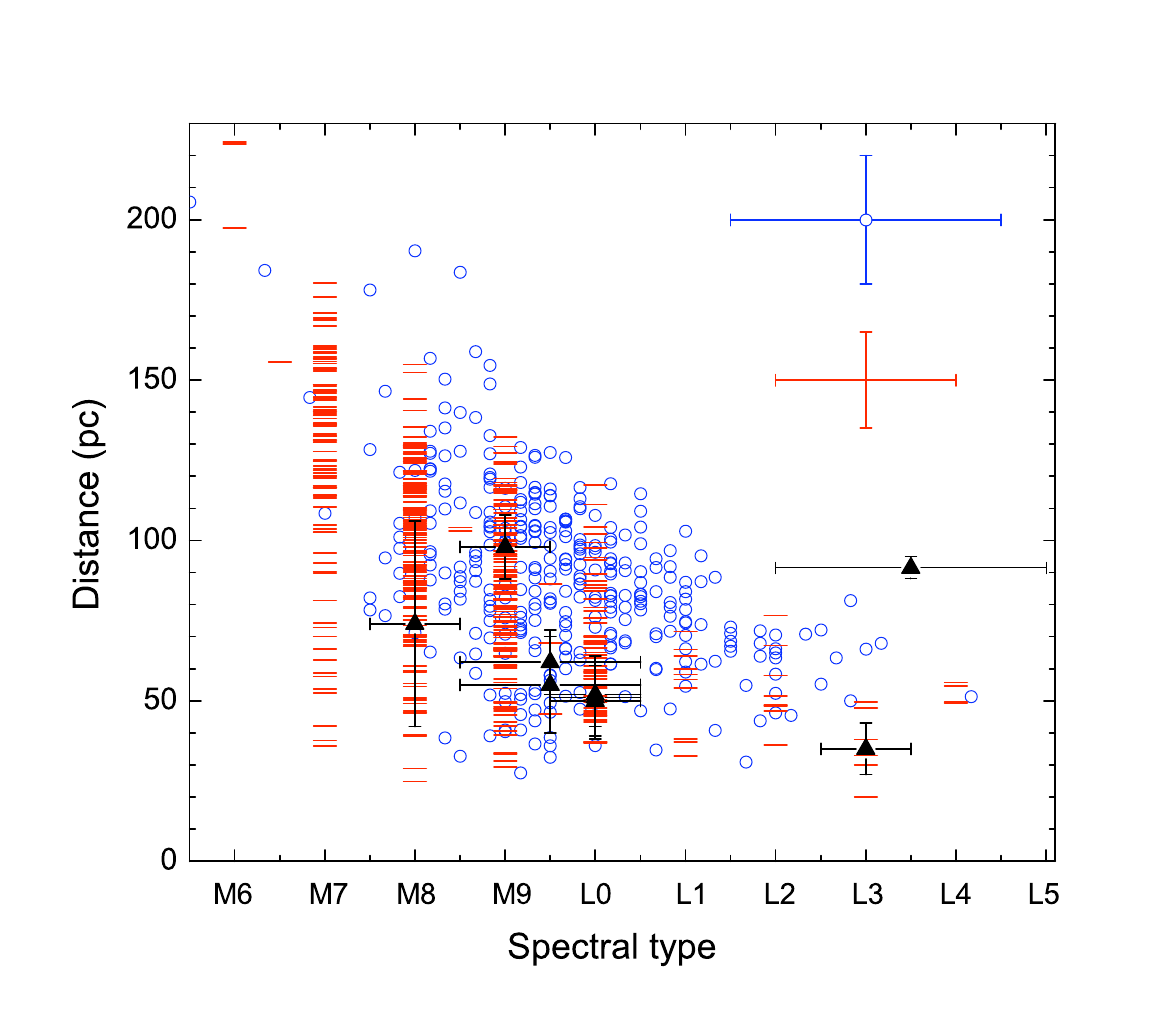}
   \caption{Plot of distance vs. spectral type for ultra-cool dwarfs identified with SDSS spectra (\emph{red bars}) and optical-infrared colours (\emph{blue circles}). Typical errors in spectral type and distance are shown in the top right. \emph{Black triangles} and the black \emph{square} (SDSS 0832) are the objects subsequently found to be in common proper motion binary systems (see Section \ref{s5.3}). For these companions we assume their distances are the same as their primary stars (see Section 5.3).}
   \label{dis}
\end{figure}

\section{Spectral types}

\subsection{SDSS red optical spectroscopy}
The SDSS imaging data are used to uniformly select different
classes of objects whose spectra will be taken with the SDSS 2.5 m
telescope \citep{yor}. The target selection
algorithms for spectroscopic follow up are described by \citet{sto}. The wavelength coverage is from 3800 to 9200 {\AA} with a resolving power
$\lambda/(\Delta\lambda)=1800$. The signal-to-noise ratio is better
than 4 pixel$^{-1}$ at $g=20.2$ \citep{ade}. SDSS spectra have been sky
subtracted and corrected for telluric absorption. The spectroscopic
data are automatically reduced by the SDSS pipeline software. For 469 ultra-cool dwarf candidates with SDSS spectra, we assigned their spectral type with the {\scriptsize HAMMER} pipeline \citep{cov}. The typical spectral type error is $\sim$0.5$-$1.0 subtypes. A comparison between our analysis and the literature \citep{rei08,wes,cru,cru03,ken07,haw,sch02,kir00} shows generally good agreement, with a small number of minor ambiguities. These are 79 L dwarfs with SDSS spectra in our sample, 11 objects are known L dwarfs (one of them is in binary system), 42 objects are catalogued by \citet{wes} as M9 or L0 dwarfs and 34 new L dwarfs are presented. Table \ref{t34ld} shows the SDSS, 2MASS and UKIDSS photometric data of the 34 new L dwarfs. Figure \ref{s39} shows the SDSS spectra of 39 ultra-cool dwarfs. SDSS name and spectral types are labeled. The six ultra-cool dwarfs that we subsequently found to be wide binary companions (see Section \ref{s5.3}) are labeled with both their original name and their new binary name.

Of these companion ultra-cool dwarfs, SDSS J020735.59+135556.2 (SDSS 0207) was characterized by \citet{haw} as an L3 dwarf using its SDSS spectrum, and SDSS J085836.97+271050.8 (SDSS 0858), SDSS J095613.13+014514.3 (SDSS 0956), SDSS 163817.31+321144.1 (SDSS 1638) and SDSS 163126.15+294836.9 (SDSS 1631) were catalogued by \citet{wes} as an L0, L0, M9 and M8 respectively. The spectra of these previously identified ultra-cool dwarfs are shown in blue. The spectra of SDSS 130433.16+090706.9 (SDSS 1304), a new L0 dwarf companion, and the other 33 new L dwarfs discovered in this work are shown in black.

\begin{landscape}
\begin{table}
 \centering
 %\begin{minipage}{140mm}
  \caption{Photometric data of 34 new L dwarfs.}
  \begin{tabular}{c c c c c c c c c c c }
  \hline\hline
Name & SDSS \emph{i} & SDSS \emph{z} & 2MASS \emph{J} & 2MASS \emph{H} & 2MASS \emph{K} & UKIDSS \emph{Y} & UKIDSS \emph{J} & UKIDSS \emph{H} & UKIDSS \emph{K} & Sp.Type \\
\hline
SDSS J000614.06+160454.5 & 20.74$\pm$0.06 & 18.92$\pm$0.05 & 16.59$\pm$0.13 & 15.84$\pm$0.11 & 15.09$\pm$0.16 & ... & ... & 15.88$\pm$0.02 & 15.32$\pm$0.02 & L0 \\
SDSS J004058.92+152845.0 & 20.59$\pm$0.06 & 18.85$\pm$0.04 & 16.96$\pm$0.21 & 15.58$\pm$0.15 & 15.60$\pm$0.23 & ... & ... & 15.96$\pm$0.02 & 15.48$\pm$0.02 & L0 \\
SDSS J084333.28+102443.5 & 19.30$\pm$0.02 & 17.46$\pm$0.01 & 14.87$\pm$0.04 & 14.09$\pm$0.04 & 13.67$\pm$0.04 & 15.99$\pm$0.01 & 14.78$\pm$0.01 & 14.10$\pm$0.01 & 13.55$\pm$0.01 & L2 \\
SDSS J084407.00+284702.1 & 20.17$\pm$0.04 & 18.25$\pm$0.03 & 15.87$\pm$0.06 & 14.82$\pm$0.05 & 14.34$\pm$0.06 & ... & 15.76$\pm$0.01 & ... & ... & L2 \\
SDSS J084610.73+030837.5 & 20.29$\pm$0.04 & 18.39$\pm$0.03 & 15.91$\pm$0.09 & 15.17$\pm$0.10 & 15.03$\pm$0.17 & 16.95$\pm$0.01 & ... & 15.29$\pm$0.01 & 14.76$\pm$0.01 & L0 \\
SDSS J092142.51+084202.7 & 19.56$\pm$0.04 & 17.82$\pm$0.03 & 15.79$\pm$0.08 & 14.89$\pm$0.09 & 14.34$\pm$0.07 & 16.80$\pm$0.01 & 15.65$\pm$0.01 & 15.00$\pm$0.01 & ... & L1 \\
SDSS J093215.45+345624.8&20.60$\pm$0.04&18.81$\pm$0.03&16.73$\pm$0.15&15.70$\pm$0.14&14.98$\pm$0.14&... &16.34$\pm$0.01&... &... &L2\\
SDSS J093242.84+042215.2 & 19.55$\pm$0.02 & 17.87$\pm$0.02 & 15.66$\pm$0.08 & 14.91$\pm$0.07 & 14.62$\pm$0.09 & ... & ... & 15.00$\pm$0.01 & 14.51$\pm$0.01 & L0 \\
SDSS J093600.12+043147.9 & 20.86$\pm$0.07 & 18.98$\pm$0.04 & 16.12$\pm$0.09 & 15.13$\pm$0.06 & 14.48$\pm$0.09 & ... & ... & 15.20$\pm$0.01 & 14.44$\pm$0.01 & L3 \\
SDSS J093858.87+044343.9 & 19.23$\pm$0.02 & 17.50$\pm$0.02 & 15.24$\pm$0.05 & 14.50$\pm$0.05 & 14.00$\pm$0.07 & ... & ... & 14.59$\pm$0.01 & 14.02$\pm$0.01 & L0 \\
SDSS J094140.65$-$003215.8 & 20.60$\pm$0.05 & 18.79$\pm$0.04 & 16.77$\pm$0.12 & 15.99$\pm$0.14 & 15.59$\pm$0.21 & 17.49$\pm$0.02 & 16.69$\pm$0.02 & 16.30$\pm$0.02 & 15.86$\pm$0.02 & L0 \\
SDSS J101304.34+071050.7 & 20.51$\pm$0.04 & 18.62$\pm$0.03 & 16.22$\pm$0.14 & 15.47$\pm$0.14 & 15.09$\pm$0.18 & 17.22$\pm$0.02 & 16.19$\pm$0.01 & 15.60$\pm$0.01 & 15.07$\pm$0.01 & L0 \\
SDSS J121304.77+152922.2&20.69$\pm$0.07&18.76$\pm$0.05&16.33$\pm$0.11&15.66$\pm$0.12&15.42$\pm$0.16&... &... &15.69$\pm$0.02&15.18$\pm$0.02&L4\\
SDSS J121917.86+151612.4 & 19.46$\pm$0.02 & 17.88$\pm$0.02 & 15.68$\pm$0.07 & 15.09$\pm$0.08 & 14.71$\pm$0.11 & ... & ... & 15.08$\pm$0.01 & 14.62$\pm$0.01 & L0 \\
SDSS J124514.95+120442.0 & 20.53$\pm$0.04 & 18.67$\pm$0.03 & 16.05$\pm$0.08 & 15.32$\pm$0.11 & 14.73$\pm$0.09 & 17.14$\pm$0.02 & 16.06$\pm$0.01 & 15.33$\pm$0.01 & 14.67$\pm$0.01 & L1 \\
SDSS J125214.08+142239.3 & 20.56$\pm$0.04 & 18.76$\pm$0.03 & 16.00$\pm$0.07 & 15.16$\pm$0.08 & 14.49$\pm$0.07 & 17.28$\pm$0.02 & 16.05$\pm$0.01 & 15.24$\pm$0.01 & 14.58$\pm$0.01 & L3 \\
SDSS J125331.83+065933.4 & 20.20$\pm$0.04 & 18.35$\pm$0.03 & 16.19$\pm$0.11 & 15.27$\pm$0.10 & 15.07$\pm$0.16 & 17.01$\pm$0.01 & 16.07$\pm$0.01 & 15.48$\pm$0.01 & 14.96$\pm$0.01 & L0 \\
~~SDSS J130433.16+090706.9$^{a}$ & 19.11$\pm$0.02 & 17.33$\pm$0.02 & 15.29$\pm$0.06 & 14.57$\pm$0.07 & 13.95$\pm$0.07 & 16.32$\pm$0.01 & 15.28$\pm$0.01 & 14.59$\pm$0.01 & 14.03$\pm$0.01 & L0 \\
SDSS J130831.02+081852.3 & 19.60$\pm$0.02 & 17.82$\pm$0.02 & 15.13$\pm$0.05 & 14.35$\pm$0.06 & 13.85$\pm$0.07 & 16.31$\pm$0.01 & 15.19$\pm$0.01 & 14.37$\pm$0.01 & 13.79$\pm$0.01 & L0 \\
SDSS J132715.21+075937.5 & 19.17$\pm$0.02 & 17.33$\pm$0.01 & 14.60$\pm$0.04 & 13.79$\pm$0.04 & 13.24$\pm$0.04 & 15.75$\pm$0.01 & 14.58$\pm$0.01 & 13.83$\pm$0.01 & 13.20$\pm$0.01 & L1 \\
SDSS J134607.41+084234.5 & 20.04$\pm$0.03 & 18.16$\pm$0.02 & 15.74$\pm$0.07 & 14.79$\pm$0.08 & 14.16$\pm$0.07 & 16.74$\pm$0.01 & 15.52$\pm$0.01 & 14.75$\pm$0.01 & 14.11$\pm$0.01 & L3 \\
SDSS J142257.14+082752.1 & 19.37$\pm$0.02 & 17.60$\pm$0.02 & 15.10$\pm$0.05 & 14.22$\pm$0.03 & 13.65$\pm$0.05 & 16.24$\pm$0.01 & 15.01$\pm$0.01 & 14.27$\pm$0.01 & 13.61$\pm$0.01 & L3 \\
SDSS J143911.87+082315.6 & 19.84$\pm$0.03 & 17.93$\pm$0.02 & 15.39$\pm$0.04 & 14.65$\pm$0.05 & 14.09$\pm$0.07 & 16.54$\pm$0.01 & 15.35$\pm$0.01 & 14.69$\pm$0.01 & 14.09$\pm$0.01 & L1 \\
SDSS J144016.19+002638.9&20.43$\pm$0.05&18.57$\pm$0.03&16.07$\pm$0.11&15.41$\pm$0.12&14.82$\pm$0.14&17.21$\pm$0.02&16.01$\pm$0.01&15.28$\pm$0.01&14.69$\pm$0.01&L4 \\
SDSS J145201.33+093136.8 & 19.69$\pm$0.02 & 17.79$\pm$0.02 & 15.42$\pm$0.07 & 14.77$\pm$0.08 & 14.25$\pm$0.09 & 16.35$\pm$0.01 & 15.34$\pm$0.01 & 14.81$\pm$0.01 & 14.28$\pm$0.01 & L0 \\
SDSS J145658.17+070104.7&20.55$\pm$0.04&18.66$\pm$0.03&16.28$\pm$0.11&15.20$\pm$0.09&14.60$\pm$0.10&17.24$\pm$0.01&15.97$\pm$0.01&15.25$\pm$0.01&14.59$\pm$0.01&L4\\
SDSS J150309.53+115323.1 & 20.68$\pm$0.04 & 18.74$\pm$0.04 & 16.30$\pm$0.10 & 15.40$\pm$0.10 & 14.79$\pm$0.09 & 17.32$\pm$0.01 & 16.14$\pm$0.01 & 15.42$\pm$0.01 & 14.80$\pm$0.01 & L2 \\
SDSS J152314.46+105258.9 & 20.53$\pm$0.04 & 18.57$\pm$0.03 & 16.09$\pm$0.10 & 15.15$\pm$0.08 & 14.82$\pm$0.12 & 17.08$\pm$0.01 & 15.94$\pm$0.01 & 15.20$\pm$0.01 & 14.62$\pm$0.01 & L0 \\
SDSS J162307.37+290827.6&20.38$\pm$0.05&18.50$\pm$0.04&16.08$\pm$0.09&15.50$\pm$0.11&14.97$\pm$0.10&... &16.13$\pm$0.01&... &... &L4\\
SDSS J163748.64+275254.6 & 19.01$\pm$0.02 & 17.29$\pm$0.01 & 14.95$\pm$0.04 & 14.16$\pm$0.05 & 13.84$\pm$0.05 & ... & 14.89$\pm$0.01 & ... & ... & L0 \\
SDSS J164911.16+300048.3 & 20.26$\pm$0.03 & 18.38$\pm$0.03 & 16.12$\pm$0.09 & 15.38$\pm$0.11 & 14.92$\pm$0.09 & ... & 16.09$\pm$0.01 & ... & ... & L1 \\
SDSS J220517.48$-$003710.3&21.02$\pm$0.06&19.13$\pm$0.05&16.82$\pm$0.14&16.24$\pm$0.20&15.46$\pm$0.19&17.79$\pm$0.03&... &15.98$\pm$0.02&15.52$\pm$0.02&L0\\
SDSS J233615.99+004253.4&20.94$\pm$0.07&19.05$\pm$0.04&17.08$\pm$0.20&15.71$\pm$0.14&15.29$\pm$0.16&17.59$\pm$0.02&16.63$\pm$0.02&15.98$\pm$0.02&15.37$\pm$0.02&L0\\
SDSS J234759.77$-$001546.9&20.86$\pm$0.07&19.10$\pm$0.06&16.40$\pm$0.13&16.15$\pm$0.20&15.46$\pm$&17.77$\pm$0.02&16.90$\pm$0.02&16.31$\pm$0.03&15.88$\pm$0.03&L0\\
\hline
\end{tabular}
%\end{minipage}
\begin{list}{}{}
\item
Note: SDSS magnitude limits (95\% detection repeatability for point
sources) for \emph{i} and \emph{z} bands are 21.3
and 20.5 respectively; UKIDSS magnitude limits for \emph{Y}, \emph{J}, \emph{H} and
\emph{K} bands are 20.5, 20.0, 18.8 and 18.4 respectively. \\
\item[$^{a}$] Companion to an M4.5 dwarf, SDSS J130432.93+090713.7, see Table \ref{t8bi}. \\
\end{list}
%\end{center}
\label{t34ld}
\end{table}
\end{landscape}

\subsection{Spectral types from colours}
We used optical$-$near-infrared colour$-$spectral type relationships \citep{zh09a,zh09b} to assign the spectral types of 337 objects without SDSS spectra. More specifically, we used three spectral type$-$colour (\emph{i-J, i-H, i-K}) relationships, and took the average value of the three results. The typical error of the spectral type is about $\pm$1.5 subtypes. Figure \ref{dis} shows the spectral types of the ultra-cool dwarfs from both typing methods (as well as their distance estimates; see Section 4).

\subsection{Near-infrared spectral type}

\begin{figure*} %  figure placement: here, top, bottom, or page
   \centering
   \includegraphics[width=400pt]{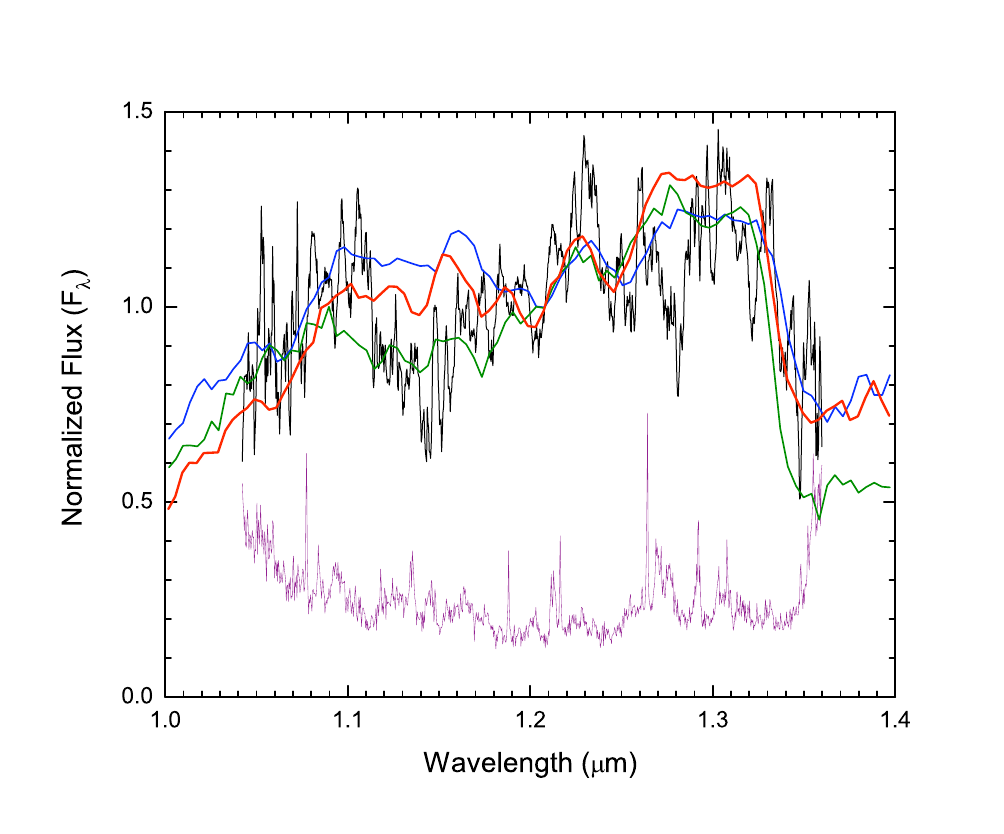}
   \caption{Gemini/NIRI spectrum of SDSS 0832. The spectrum (\emph{black}) is normalized to one at 1.208 $\mu$m and smoothed by 11 pixels. The error spectrum is plotted in \emph {purple}. 
  Three standard near-infrared spectra are over plotted, 2MASP J0345432+254023 \citep[\emph{blue}, L0 in optical, L1 in near-infrared;][]{bur06,kir99}, 2MASSW J1146345+223053 \citep[\emph{red}, L3;][]{bur10,kir99,kna04}, and DENIS-P J0205.4-1159 \citep[\emph{green}, L7 in optical, L5.5 in near-infrared;][]{bur10,kir99}.}
   \label{nir}
\end{figure*}

Spectroscopy in the $J$-band was obtained for ULAS~0832 using the
Near InfraRed Imager
and Spectrometer \citep[NIRI;][]{hod} on the
Gemini North Telescope on Mauna Kea (under program GN-2009A-Q-16) on the
8th of April 2009 through thin cirrus cloud.
All observations were made up of a set of sub-exposures in an ABBA
jitter pattern to facilitate effective background subtraction, with a
slit width of 1\arcsec.
The length of the A-B jitter was 10\arcsec, and the integration
consisted of 6 sub-exposures of 240~seconds.

The NIRI observations were reduced using standard {\scriptsize IRAF} Gemini packages.
A comparison argon arc frame was
used to obtain a dispersion solution, which was applied to the
pixel coordinates in the dispersion direction on the images.
The resulting wavelength-calibrated subtracted AB-pairs had a low-level
of residual sky emission removed by fitting and subtracting this
emission with a set of polynomial functions fit to each pixel row
perpendicular to the dispersion direction, and considering pixel data
on either side of the target spectrum only.
The spectra were then extracted using a linear aperture, and cosmic
rays and bad pixels removed using a sigma-clipping algorithm.

Telluric correction was achieved by dividing the extracted target
spectra by that of the F6V star HIP 41187, observed just before the
target.
Prior to division, hydrogen lines were removed from the standard star
spectrum by interpolating the stellar
continuum.
Relative flux calibration was then achieved by multiplying through by a
blackbody spectrum with $T_{\rm eff} = 6300$K.

Figure \ref{nir} shows this spectrum and three standard spectra (L0, L3 and L7).  The spectrum is somewhat noisy, but it but is consistent with early L spectral type by the near-infrared characterization scheme of \citet{geb}.

\section{Distances}
 \label{s4}
We used the M$_{J}-$spectral type relationship and M$_{i}- (i-J)$ relationship of \citet{haw} to estimate distances of ultra-cool dwarfs with and without SDSS spectra respectively. We have assumed two limiting cases. Firstly we assume that the ultra-cool dwarfs are single objects (these are the distances shown in Figure \ref{dis} and refered to as ``Dist'' in the subsequent tables). However, to provide conservative upper limits we also perform the calculations assuming that each candidate is an unresolved equal-mass binary giving distances 41\% larger (referred to as Dist.b in subsequent tables). In each case the uncertainties were calculated to allow for spectral type uncertainty and the RMS scatter in the absolute magnitude$-$spectral type/colour relationships.

\section{Proper motions}

In previous work by our group we have measured proper motions using various combinations of survey databases \citep{zh09b} and found that 2MASS and UKIDSS generally provide the most accurate proper motions due to longer baselines (5$-$10 years). The combinations of SDSS-2MASS and SDSS-UKIDSS only sometimes provide useful proper motions, when for example the baselines involved are relatively long ( $^{>}_{\sim}$ 5 years) or the measured objects have large proper motions \citep{dea,she, zh09b}.

\subsection{Ultra-cool dwarfs from our SDSS/2MASS/UKIDSS sample}
\begin{figure}%  figure placement: here, top, bottom, or page
   \centering
   \includegraphics[width=260pt]{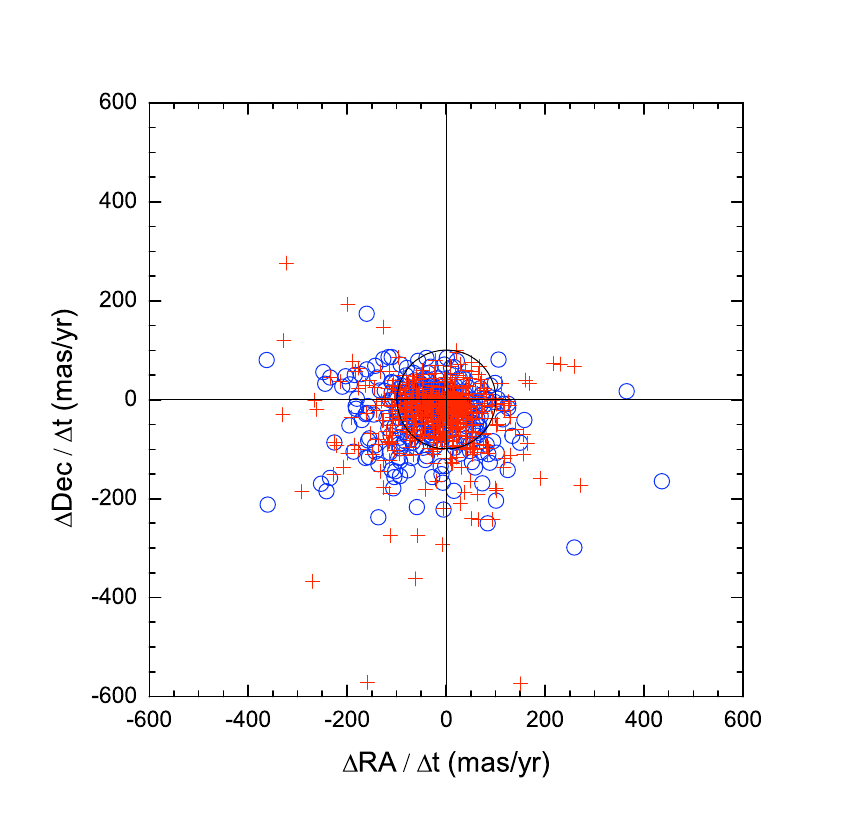}
   \caption{ 2MASS-UKIDSS proper motions of ultra-cool dwarfs identified with (\emph{red crosses}) and without (\emph{blue circles}) SDSS spectra. A ring indicates the 100mas/yr proper motion locus, which is a useful division when searching for common proper motion companions (see text).}
   \label{pms}
\end{figure}

806 of our objects matched in 2MASS and UKIDSS, so we calculated proper motions from their database coordinates and epochs, following the method described in \citet[Section 3;][]{zh09b}. Figure \ref{pms} shows the resulting 2MASS-UKIDSS proper motions of these 806 ultra-cool dwarfs. The 100mas/yr ring shown in the Figure \ref{pms} acts as a guide to identify ultra-cool dwarfs whose proper motions are comparable with those in existing high proper motion catalogues ($>$100mas/yr). This separation is useful when searching for possible companions (see Section 5.3). Table \ref{t34pm} shows the proper motions and distance estimates for the 34 new L dwarfs.

\begin{figure}%  figure placement: here, top, bottom, or page
   \centering
   \includegraphics[width=\columnwidth]{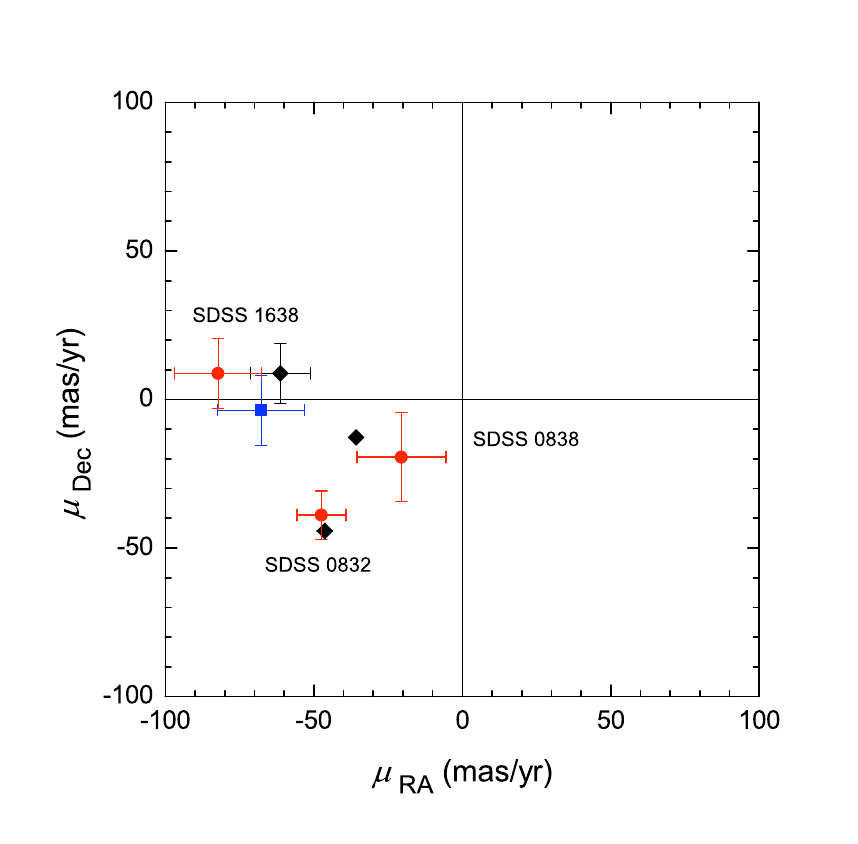}
   \caption{The proper motions of SDSS 0832 (L3.5), SDSS 0838 (M9.5), SDSS1638 (L0) and their primary star or cluster (\emph{squares} and a \emph{diamond} respectively). The proper motions of these three ultra-cool dwarfs (\emph{red circles}) and the primary star of SDSS 1638 (\emph{blue sqrare}) are measured with the {\scriptsize IRAF} routines {\scriptsize GEOMAP} and {\scriptsize GEOXYTRAN}. The \emph{black diamonds} are proper motions from the literature. SDSS 0832 is the companion to the K3 giant $\eta$ Cancri. SDSS 0838 is a Praesepe candidate member. SDSS 1638 is a companion to an M4 dwarf.}
   \label{pm3}
\end{figure}

\begin{figure} %  figure placement: here, top, bottom, or page
   \centering
   \includegraphics[width=\columnwidth]{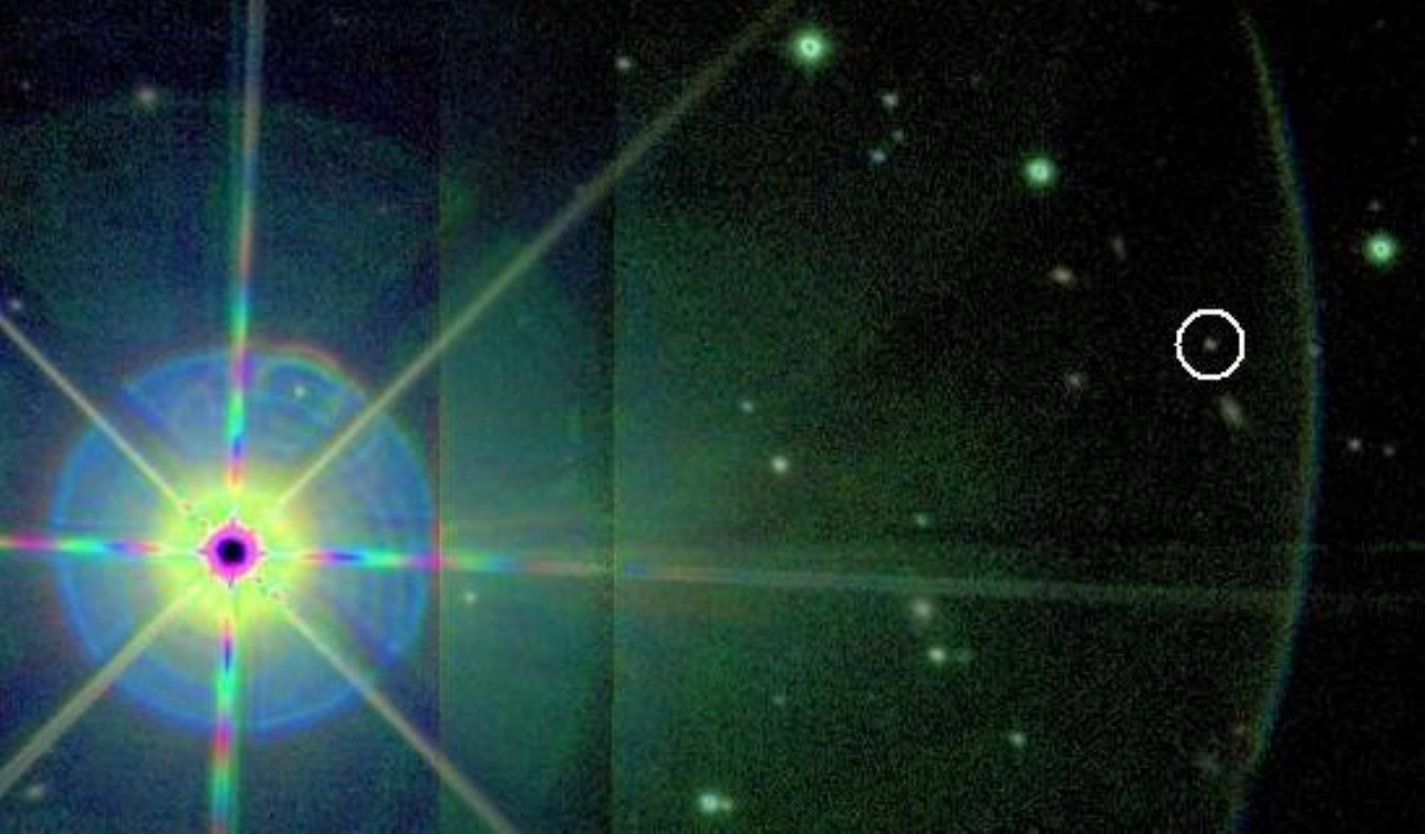}
   \caption{A UKIDSS \emph{JHK} false colour image of $\eta$ Cancri AB with a separation of 164 arcsec.  $\eta$ Cancri A is the brightest object in this field, $\eta$ Cancri B is in the white circle.}
\label{eta}
\end{figure}

\begin{figure} %  figure placement: here, top, bottom, or page
   \centering
   \includegraphics[width=\columnwidth]{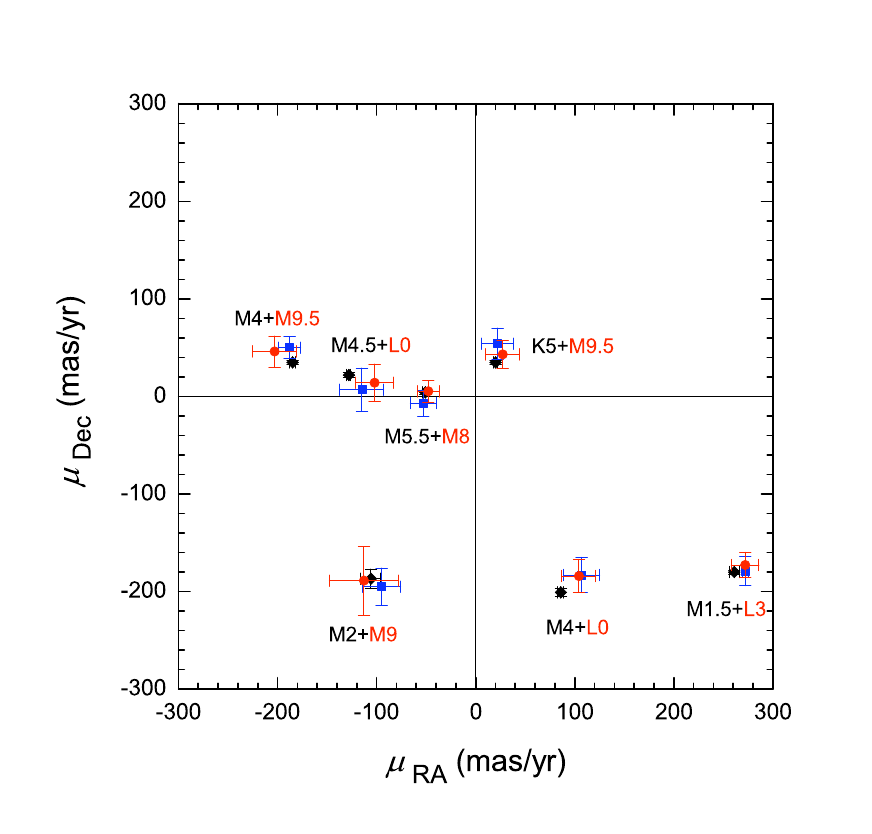}
   \caption{Seven dwarf star$+$ultra-cool dwarf common proper motion pairs. Proper motions of ultra-cool dwarfs (\emph{red circles}) and dwarf stars (\emph{blue squares}) are measured in this work using 2MASS and UKIDSS database coordinates and epochs. Dwarf star proper motions taken from the literature (\emph{black diamonds}) are also shown.}
   \label{pm7}
\end{figure}

\subsection{Ultra-cool dwarfs from our deeper search towards Praesepe}

Due to the relatively short baselines between the SDSS and UKIDSS observations of our ultra-cool dwarfs towards Praesepe, and their low optical signal-to-noise, we obtained additional epoch imaging for some of our candidates. \emph{z}-band images were obtained for two candidates, SDSS J083823.69$+$164702.0 (SDSS 0838) and SDSS J083231.78+202700.0 (SDSS 0832), using the ESO Faint Object Spectrograph and Camera (v.2) \citep[EFOSC2;][]{buz} on the New Technology Telescope (NTT) on La Silla, ESO (under program 082.C-0399B) on 25th December 2008 with an exposure time of 500 seconds. The EFOSC2 observations were reduced using standard {\scriptsize IRAF} packages. We then used the {\scriptsize IRAF} task {\scriptsize GEOMAP} to derive spatial transformations from the SDSS $z$ or UKIDSS GCS $K$-band images into the NTT $z$-band images. These transforms allowed for linear shifts and rotation. We then transformed the SDSS or GCS pixel coordinates of objects into the NTT images using {\scriptsize GEOXYTRAN}, and calculated the change in position (relative to the reference stars) between the two epochs.

We measured the proper motion of SDSS 0838 ( $\mu_{\rm RA}$ = $-$21$\pm$15 mas/yr, $ \mu_{\rm Dec}$ = $-$19$\pm$15 mas/yr; 2.2 year baseline) based on GCS and NTT images with a baseline of 2.2 years and using 14 reference stars. This data does not yield a proper motion that is greater than zero at a strongly significant level, although we note that the measurement is potentially consistent with the proper motion of Praesepe \citep[ $\mu_{\rm RA}$ = $-$35.81 mas/yr,  $\mu_{\rm Dec}$ = $-$12.85 mas/yr;][]{va2}. The distance constraint for SDSS 0838 (see Section \ref{s4}; 181$\pm$62 pc if single, or 256$\pm$62 pc if an unresolved equal-mass binary) is also consistent with that of the cluster \citep[181.5 pc;][]{va2}. A longer base-line is needed before cluster membership can be rigorously assessed, but this object could be a member of Praesepe. Table \ref{t0838} shows the parameters of SDSS 0838.

We measured the proper motion of SDSS 0832 ( $\mu_{\rm RA}$ = $-$46.6$\pm$8.2 mas/yr,  $\mu_{\rm Dec}$ = $-$37.0$\pm$8.2 mas/yr) based on SDSS and NTT images with a baseline of 4.7 years and using 16 reference stars. In addition, the combination of SDSS and UKIDSS GCS epochs provided a proper motion ($\mu_{\rm RA}$ = $-$52$\pm$46 mas/yr, $\mu_{\rm Dec}$ = $-$50$\pm$46 mas/yr) with a baseline of 2.6 years \citep[][]{zh09b} with significantly larger uncertainties. We combined these two measurements together, weighting by their reciprocal uncertainties, to give a final proper motion of $\mu_{\rm RA}$ = $-$47.4$\pm$8.2 mas/yr, $\mu_{\rm Dec}$ = $-$38.9$\pm$8.2 mas/yr for SDSS 0832. Both this proper motion and the distance constraint for SDSS 0832 (121$\pm$33 pc) are inconsistent with the Praesepe cluster, and this ultra-cool dwarf is thus seen to be a foreground object in the field.

\begin{table}
 \centering
 %\begin{minipage}{140mm}
  \caption{Proper motions and distance estimates for the 34 new L dwarfs.}
  \begin{tabular}{l r r c r }
  \hline\hline
Name &  $\mu_{\rm RA}$$^{a}$ & $\mu_{\rm Dec}$$^{a}$ & Distance$^{b}$ & Sp.T. \\
&  (mas$/$yr) & (mas$/$yr) & (pc) &  \\
\hline
SDSS 0006 &  $-$5$\pm$19 & $-$40$\pm$19 & ~94$\pm$29 & L0 \\
SDSS 0040 &  121$\pm$37 & $-$62$\pm$37 & 111$\pm$39 & L0 \\
SDSS 0843 & 150$\pm$10 & $-$574$\pm$10 & ~32$\pm$~9 & L2 \\
SDSS 0844 &  40$\pm$~7 & $-$98$\pm$~7 & ~51$\pm$15 & L2 \\
SDSS 0846 &  19$\pm$19 & $-$18$\pm$19 & ~69$\pm$21 & L0 \\
SDSS 0921 & $-$136$\pm$15 & $-$81$\pm$15 & ~57$\pm$17 & L1 \\
SDSS 093215 & 40$\pm$25 & $-$74$\pm$25 & ~77$\pm$15 & L2  \\
SDSS 093242 &  23$\pm$11 & 12$\pm$11 & ~61$\pm$18 & L0 \\
SDSS 0936 &  $-$49$\pm$13 & $-$38$\pm$13 & ~51$\pm$15 & L3 \\
SDSS 0938 &  $-$12$\pm$11 & $-$38$\pm$11 & ~50$\pm$15 & L0 \\
SDSS 0941 &  $-$63$\pm$46 & $-$362$\pm$46 & 102$\pm$32 & L0 \\
SDSS 1013 &  17$\pm$32 & 28$\pm$32 & ~79$\pm$25 & L0 \\
SDSS 1213 & $-$292$\pm$30 & $-$186$\pm$30 & ~49$\pm$~8 & L4  \\
SDSS 1219 &  $-$32$\pm$30 & $-$13$\pm$30 & ~62$\pm$18 & L0 \\
SDSS 1245 &  4$\pm$18 & $-$43$\pm$18 & ~64$\pm$19 & L1 \\
SDSS 1252 &  $-$66$\pm$23 & $-$50$\pm$23 & ~48$\pm$14 & L3 \\
SDSS 1253 &  $-$14$\pm$21 & $-$1$\pm$21 & ~78$\pm$24 & L0 \\
SDSS 1304 &  $-$102$\pm$19 & 14$\pm$19 & ~51$\pm$14 & L0 \\
SDSS 1308 &  $-$234$\pm$13 & 45$\pm$13 & ~48$\pm$14 & L0 \\
SDSS 1327 &  $-$162$\pm$22 & 29$\pm$22 & ~33$\pm$~9 & L1 \\
SDSS 1346 &  $-$221$\pm$19 & $-$93$\pm$19 & ~43$\pm$13 & L3 \\
SDSS 1422 &  $-$159$\pm$11 & $-$571$\pm$11 & ~32$\pm$~9 & L3 \\
SDSS 1439 &  $-$130$\pm$19 & $-$3$\pm$19 & ~47$\pm$14 & L1 \\
SDSS 1440 & $-$26$\pm$26 & 10$\pm$26 & ~43$\pm$~7 & L4  \\
SDSS 1452 &  65$\pm$19 & $-$191$\pm$19 & ~55$\pm$16 & L0 \\
SDSS 1456 & 37$\pm$17 & $-$16$\pm$17 & ~48$\pm$~7 & L4  \\
SDSS 1503 &  22$\pm$36 & 96$\pm$36 & ~63$\pm$19 & L2 \\
SDSS 1523 &  0.4$\pm$19 & $-$19$\pm$19 & ~74$\pm$22 & L0 \\
SDSS 1623 & $-$265$\pm$16 & $-$1$\pm$16 & ~43$\pm$~7 & L4  \\
SDSS 1637 &  $-$226$\pm$13 & $-$86$\pm$13 & ~44$\pm$13 & L0 \\
SDSS 1649 &  $-$27$\pm$21 & 15$\pm$21 & ~66$\pm$20 & L1 \\
SDSS 2205 & 51$\pm$33 & 75$\pm$33 & 104$\pm$27 & L0 \\
SDSS 2336 & $-$44$\pm$45 & $-$27$\pm$45 & 117$\pm$31 & L0  \\
SDSS 2347 & 13$\pm$40 & $-$6$\pm$40 & ~86$\pm$23 & L0  \\
\hline
\end{tabular}
%\end{minipage}
\begin{list}{}{}
\item[$^{a}$] 2MASS-UKIDSS data-base proper motions $-$ found by dividing the difference between
the 2MASS and SDSS coordinates (from the respective databases) by
the observational epoch difference.
Standard errors are calculated using the major axes of the position error ellipses from 2MASS and SDSS. $^{b}$ Distance are based on M$_{J}-$spectral type relationship \citep{haw}. \\
\end{list}
%\end{center}
\label{t34pm}
\end{table}

\begin{table*}
 \centering
 %\begin{minipage}{140mm}
  \caption{Proper motions of companions of eight dwarf + ultra-cool dwarf binary systems.}
  \begin{tabular}{l c c c c c c c c  }
  \hline\hline
Primary  & \multicolumn{2}{c}{TYC 1189-1216-1}   & \multicolumn{2}{c}{G 3-40}  &  \multicolumn{2}{c}{LP 312-49}   & \multicolumn{2}{c}{LP 548-50} \\
\hline
Ref.$^{a}$  & $\mu_{\rm RA}$ & $\mu_{\rm Dec}$ & $\mu_{\rm RA}$ & $\mu_{\rm Dec}$ & $\mu_{\rm RA}$ & $\mu_{\rm Dec}$ & $\mu_{\rm RA}$ & $\mu_{\rm Dec}$  \\
& (mas$/$yr) & (mas$/$yr) & (mas$/$yr) & (mas$/$yr)  & (mas$/$yr) & (mas$/$yr) & (mas$/$yr) & (mas$/$yr) \\
------------ & \multicolumn{2}{c}{--------------------------}   & \multicolumn{2}{c}{-----------------------------}  &  \multicolumn{2}{c}{-----------------------------}   & \multicolumn{2}{c}{-----------------------------} \\
$^{1}$ & ... & ... & 259 ... & $-$188 ... & ... & ... & ... & ... \\
$^{2}$ & ... & ... & ... &  ... & 113 ... & $-$196 ... & $-$184 ... & 32 ... \\
$^{3}$ &  20$\pm$3 & 35$\pm$3 & ... & ... & ... & ... & ... & ... \\
$^{4}$ & 21$\pm$3 & 35$\pm$3 & 261$\pm$15 & $-$165$\pm$15 & ... & ... & ... & ... \\
$^{5}$ & 18 ... & 34 ... & 254$\pm$4 & $-$176$\pm$1 & 80$\pm$1 & $-$198$\pm$3 & $-$182$\pm$3 & 36$\pm$2 \\
$^{6}$ & ... & ... & 261$\pm$6 & $-$185$\pm$6 & 81$\pm$6 & $-$205$\pm$6 & ... & ... \\
$^{7}$ & 20$\pm$2 & 35$\pm$2 & 263$\pm$2 & $-$180$\pm$2 & 83$\pm$4 & $-$202$\pm$5 & $-$182$\pm$3 & 36$\pm$2 \\
$^{8}$ & 20$\pm$2 & 33$\pm$2 & 262 ... & $-$186 ... & 82 ... & $-$201 ... & ... & ... \\
$^{9}$ & ... & ... & 262 ... & $-$186 ... & 82 ... & $-$201 ... & $-$186 ... & 37 ... \\
$^{10}$ & 22$\pm$1 & 37$\pm$1 & 265$\pm$2 & $-$181$\pm$2 & ... & ... & ... & ... \\
$^{11}$ & ... &  ... & 256$\pm$3 & $-$176$\pm$3 & 81$\pm$3 & $-$200$\pm$3 & $-$191$\pm$3 & 36$\pm$3 \\
$^{12}$ & 20 ... & 35 ... & 264 ... & $-$180 ... & 83 ... & $-$202 ... & ... & ... \\
$^{13}$ & 21$\pm$1 & 36$\pm$1 & 264$\pm$3 & $-$179$\pm$3 & ... & ... & ... & ... \\
$^{14}$ & 22$\pm$16 & 54$\pm$16 & 272$\pm$15 & $-$179$\pm$15 & 107$\pm$18 & $-$183$\pm$18 & $-$188$\pm$11 & 50$\pm$11 \\
------------ & \multicolumn{2}{c}{--------------------------}   & \multicolumn{2}{c}{-----------------------------}  &  \multicolumn{2}{c}{-----------------------------}   & \multicolumn{2}{c}{-----------------------------} \\
Average$^{b}$   & 20$\pm$2 & 37$\pm$2 & 262$\pm$2 & $-$179$\pm$2 & 89$\pm$6 & $-$199$\pm$3 & $-$186$\pm$3 & 38$\pm$3 \\
\hline
Secondary$^{c}$   & 27$\pm$17 & 43$\pm$14 & 273$\pm$19 & $-$174$\pm$17 & 104$\pm$20 & $-$184$\pm$20 & $-$203$\pm$17 & 46$\pm$17 \\
\hline\hline
Primary  & \multicolumn{2}{c}{LP 609-3}   & \multicolumn{2}{c}{SDSS J130432.93+090713.7}  &  \multicolumn{2}{c}{SDSS J163126.17+294847.1}   & \multicolumn{2}{c}{SDSS J163814.32+321133.5} \\
\hline
Ref.$^{a}$  & $\mu_{\rm RA}$ & $\mu_{\rm Dec}$ & $\mu_{\rm RA}$ & $\mu_{\rm Dec}$ & $\mu_{\rm RA}$ & $\mu_{\rm Dec}$ & $\mu_{\rm RA}$ & $\mu_{\rm Dec}$  \\
& (mas$/$yr) & (mas$/$yr) & (mas$/$yr) & (mas$/$yr)  & (mas$/$yr) & (mas$/$yr) & (mas$/$yr) & (mas$/$yr) \\
------------ & \multicolumn{2}{c}{--------------------------}   & \multicolumn{2}{c}{-----------------------------}  &  \multicolumn{2}{c}{-----------------------------}   & \multicolumn{2}{c}{-----------------------------} \\
$^{2}$ & $-$105 & $-$182 & ... &  ... & ... &  ... &  ... &  ... \\
$^{5}$ & ... & ... & $-$126$\pm$1 & 22$\pm$3 & $-$52$\pm$20 & 6$\pm$1 & $-$60$\pm$1 & 2 ... \\
$^{7}$ & ... & ... & $-$126$\pm$1 & 22$\pm$3 & $-$52$\pm$20 & 6$\pm$1 & $-$60$\pm$1 & 2 ... \\
$^{8}$ & $-$107 ... & $-$189 ... & ... & ... & ... & ... & ... & ... \\
$^{9}$ & $-$107 ... & $-$189 ... & ... & ... & ... & ... & ... & ... \\
$^{11}$ & ... & ... & $-$132$\pm$3 & 22$\pm$3 & $-$49$\pm$3 & 1$\pm$3 & $-$66$\pm$3 & 3$\pm$3 \\
$^{14}$ & $-$95$\pm$19 & $-$195$\pm$19 & $-$115$\pm$22 & 7$\pm$22 & $-$53$\pm$13 & $-$7$\pm$13 & $-$85$\pm$10 & 2$\pm$10 \\
$^{15}$ & ...  & ... & ... & ... & ... & ... & $-$68$\pm$15 & $-$4$\pm$12 \\
------------ & \multicolumn{2}{c}{--------------------------}   & \multicolumn{2}{c}{-----------------------------}  &  \multicolumn{2}{c}{-----------------------------}   & \multicolumn{2}{c}{-----------------------------} \\
Average$^{b}$   & $-$104$\pm$14 & $-$189$\pm$19 & $-$125$\pm$3 & 18$\pm$4 & $-$52$\pm$3 & 2$\pm$4 & $-$66$\pm$6 & 6$\pm$5 \\
\hline
Secondary$^{c}$  & $-$113$\pm$37 & $-$190$\pm$36 & $-$102$\pm$19 & 14$\pm$19 & $-$48$\pm$11 & 5$\pm$11 & $-$82$\pm$15 & 9$\pm$12 \\

\hline
\end{tabular}
%\end{minipage}
\begin{list}{}{}
\item
%Note: The unit for all proper motions is (mas$/$yr).
%\item[$^{a}$] NLTT 7059.
\item[$^{a}$] Reference: $^{1}$\citet{gic}, $^{2}$\citet{luy}, $^{3}$\citet{hog00}, $^{4}$\citet{kh1}, $^{5}$\citet{mon}, $^{6}$\citet{sal}, $^{7}$\citet{zac04}, $^{8}$\citet{zac09}, $^{9}$\citet{lep}, $^{10}$\citet{duc}, $^{11}$\citet{ade}, $^{12}$\citet{iva}, $^{13}$\citet{ros}, $^{14}$measured with database information in this work, $^{15}$measured with {\scriptsize IRAF} in this work.
\item[$^{b}$] Average proper motions of primary stars.
\item[$^{c}$] Proper motions of secondary ultra-cool dwarfs.
\end{list}
%\end{center}
\label{t8pm}
\end{table*}

\begin{table*}
 \centering
 %\begin{minipage}{140mm}
  \caption{Parameters of eight ultra-cool dwarf binary systems.}
  \begin{tabular}{l c c c c c c c c  }
  \hline\hline
Binary & A & B & A & B & A & B & A & B  \\
Component & TYC &  SDSS 0101   & G 3-40 &  SDSS 0207   & LP 312-49 &  SDSS 0858   &   LP 548-50 &  SDSS 0953  \\
&1189-1216-1 \\
 ---------------  & \multicolumn{2}{c}{-----------------------------------}   & \multicolumn{2}{c}{-----------------------------------}  &  \multicolumn{2}{c}{-----------------------------------}   & \multicolumn{2}{c}{-----------------------------------} \\
Sp. Type & K5 & M9.5$^{g}$  & M1.5 & L3$^{c}$ & M4 & L0$^{d}$ &M4 & M9.5$^{g}$ \\
SDSS RA & 01 01 55.00 & 01 01 53.11  & 02 07 37.47 & 02 07 35.59  & 08 58 36.75 & 08 58 36.97 &  09 53 24.20 & 09 53 24.54 \\
SDSS Dec &  15 28 01.2 & 15 28 19.4 & 13 54 49.4~ & 13 55 56.2~  & 27 11 05.8~ & 27 10 50.8~ & 05 27 00.9~ & 05 26 58.4~\\
Dist (pc) & 55$\pm$15 & 60$\pm$18   & 35$^{e}$$\pm$8 & 37$\pm$10  & 52$\pm$10 & 46$\pm$11 & 62$\pm$10 & 65$\pm$13 \\
Dist.b$^{a}$(pc) & ... & 85$\pm$18  & ...   & 53$\pm$10 & ... & 65$\pm$11  & ...  & 92$\pm$13 \\
 $\mu_{\rm RA}$(mas$/$yr) &  20$\pm$2 & 27$\pm$17 & ~~262$\pm$2 & ~~272$\pm$14 &  ~~88$\pm$6 & ~~104$\pm$17  & $-$186$\pm$3~~ & $-$203$\pm$22~~  \\
 $\mu_{\rm Dec}$(mas$/$yr) & 37$\pm$2 & 43$\pm$14  & $-$179$\pm$2 & $-$173$\pm$13 & $-$199$\pm$3~~ & $-$184$\pm$17 & ~~38$\pm$3 & ~~47$\pm$16 \\
%$[$Fe/H$]$$^{h}$ & ... &... & 0.13$\pm$0.32 &... & $-$0.15$\pm$0.26 &... & $-$0.17$\pm$0.21 &... \\
$V$ & 11.49$\pm$0.15 & ... & 12.50$\pm$0.13 &...&16.19&...&17.96& ... \\
SDSS $i$ & 12.79$\pm$0.01 & 19.57$\pm$0.03  & 11.01$\pm$0.01 &19.75$\pm$0.03 & 14.34$\pm$0.01 & 19.44$\pm$0.02 & 15.64$\pm$0.01 & 19.70$\pm$0.04 \\
SDSS $z$ & 10.95$\pm$0.01 & 17.74$\pm$0.02 & 11.24$\pm$0.01 &17.99$\pm$0.02 & 13.49$\pm$0.01 &17.64$\pm$0.02 & 14.69$\pm$0.01 & 17.83$\pm$0.03 \\
UKIDSS $Y$&  ...$^{f}$ & 16.31$\pm$0.01 &  ...$^{f}$ & 16.59$\pm$0.01 & ... & ... & ... & ... \\
UKIDSS $J$ &  ...$^{f}$ & 15.31$\pm$0.01 &  ...$^{f}$ & 15.37$\pm$0.01 & 11.92$\pm$0.01 & 15.01$\pm$0.01 & 13.12$\pm$0.01 & ... \\
UKIDSS $H$ &  ...$^{f}$ & 14.68$\pm$0.01 & ...$^{f}$ & 14.53$\pm$0.01 & ...  &  ... & 12.58$\pm$0.01 & 15.05$\pm$0.01\\
UKIDSS $K$  &  ...$^{f}$ & 14.15$\pm$0.01 & ...$^{f}$ & 13.83$\pm$0.01  &  ... &  ... & 12.24$\pm$0.01 & 14.47$\pm$0.01 \\
2MASS $J$ &  9.10$\pm$0.02 & 15.49$\pm$0.05 & 9.20$\pm$0.02 & 15.46$\pm$0.05  & 11.99$\pm$0.02 & 15.05$\pm$0.05 & 13.18$\pm$0.03 & 15.67$\pm$0.09 \\
2MASS $H$ & 8.48$\pm$0.03 & 14.70$\pm$0.06 & 8.57$\pm$0.06 & 14.47$\pm$0.04 & 11.41$\pm$0.02 & 14.23$\pm$0.05 & 12.57$\pm$0.03 & 14.96$\pm$0.07  \\
2MASS $K$ &  8.34$\pm$0.02 & 14.21$\pm$0.06 & 8.31$\pm$0.02 & 13.81$\pm$0.05  & 11.11$\pm$0.02 & 13.66$\pm$0.05 & 12.29$\pm$0.03 & 14.39$\pm$0.08 \\
  & \multicolumn{2}{c}{-----------------------------------}   & \multicolumn{2}{c}{-----------------------------------}  &  \multicolumn{2}{c}{-----------------------------------}   & \multicolumn{2}{c}{-----------------------------------} \\
   Separation  & \multicolumn{2}{c}{36$^{\prime\prime}$.6} & \multicolumn{2}{c}{72$^{\prime\prime}$.5} & \multicolumn{2}{c}{15$^{\prime\prime}$.4} & \multicolumn{2}{c}{5$^{\prime\prime}$.7} \\
 Sep. (AU) & \multicolumn{2}{c}{2050$\pm$571}& \multicolumn{2}{c}{2538$\pm$580}& \multicolumn{2}{c}{801$\pm$154} & \multicolumn{2}{c}{352$\pm$57} \\
 SP$^{\mathrm{b}}$ & \multicolumn{2}{c}{4.82$\times10^{-3}$} & \multicolumn{2}{c}{1.75$\times10^{-3}$} & \multicolumn{2}{c}{6.60$\times10^{-5}$} & \multicolumn{2}{c}{1.62$\times10^{-5}$} \\
\hline\hline
Binary & A & B & A & B & A & B & A & B \\
Component & LP 609-3 &   SDSS 0956 &  SDSS & SDSS  & SDSS  & SDSS  & SDSS & SDSS  \\
& & & 130432$^{h}$ & 130433$^{h}$ &  163126.17$^{h}$ &  163126.15$^{h}$ & 163814$^{h}$ & 163817$^{h}$ \\
--------------- & \multicolumn{2}{c}{-----------------------------------}   & \multicolumn{2}{c}{-----------------------------------}  &  \multicolumn{2}{c}{-----------------------------------}   & \multicolumn{2}{c}{-----------------------------------} \\
Sp. Type & M2 & M9$^{d}$ &  M4.5 & L0 & M5.5 & M8$^{d}$  & M4 & L0$^{d}$ \\
SDSS RA & 09 56 14.86 & 09 56 13.13 &  13 04 32.93 & 13 04 33.16 & 16 31 26.17 & 16 31 26.15 & 16 38 14.32 & 16 38 17.31 \\
SDSS Dec & 01 44 58.7~ & 01 45 14.3~ & 09 07 13.7~ & 09 07 06.9~ & 29 48 47.1 & 29 48 36.9 & 32 11 33.5 & 32 11 44.1 \\
Dist (pc) & 98$\pm$10 & ~91$\pm$28 &  50$\pm$11 & 51$\pm$14 & 74$\pm$32 & ~75$\pm$20 & 51$\pm$13 &54$\pm$14 \\
Dist.b$^{a}$(pc) & ...  & 128$\pm$28 & ...  & 72$\pm$14  & ... & 106$\pm$20 & ... & 76$\pm$14 \\
 $\mu_{\rm RA}$(mas$/$yr) & $-$104$\pm$14 & $-$113$\pm$35 & $-$125$\pm$3~~ & $-$102$\pm$19~~  & $-$52$\pm$3~~ & $-$48$\pm$11~~ & $-$66$\pm$~~6 & $-$82$\pm$15~~ \\
 $\mu_{\rm Dec}$(mas$/$yr) & $-$189$\pm$19 & $-$189$\pm$35 &  ~~18$\pm$4 & ~~14$\pm$19 & ~~2$\pm$4 & ~~5$\pm$11 & ~~~6$\pm$~~3 & ~~~~~9$\pm$12~~ \\
%$[$Fe/H$]$$^{h}$ & 0.01$\pm$0.13 &... & 0.13$\pm$0.30 &... &... &...  & 0.05$\pm$0.36 &...  \\
$V$ &15.99& ... & 16.63$\pm$0.34 &...&  ... & ... &14.89& ... \\
SDSS $i$ & 14.12$\pm$0.01 & 20.30$\pm$0.04  &  14.45$\pm$0.01 & 19.11$\pm$0.02 & 17.07$\pm$0.01 & 18.87$\pm$0.01   &  13.39$\pm$0.01 & 19.81$\pm$0.03  \\
SDSS $z$ &  13.50$\pm$0.01 & 18.51$\pm$0.04  &   13.52$\pm$0.01 & 17.33$\pm$0.02 & 15.86$\pm$0.01 & 17.34$\pm$0.02  &  12.37$\pm$0.01 & 17.92$\pm$0.02  \\
UKIDSS $Y$& 12.62$\pm$0.01 & 17.20$\pm$0.02 &  12.50$\pm$0.01& 16.32$\pm$0.01  & ... & ...   & ... & ... \\
UKIDSS $J$ & 12.12$\pm$0.01 & 16.36$\pm$0.01 & 11.91$\pm$0.01 &  15.28$\pm$0.01  & 14.08$\pm$0.01 & 15.39$\pm$0.01 & 11.00$\pm$0.01 & 15.35$\pm$0.01 \\
UKIDSS $H$ & 11.66$\pm$0.01 & 15.92$\pm$0.01 &  11.41$\pm$0.01 & 14.59$\pm$0.01 & ... & ...  & ... & ...\\
UKIDSS $K$  & 11.40$\pm$0.01 & 15.48$\pm$0.02 &  11.10$\pm$0.01 & 14.03$\pm$0.01 & ... &  ... & ... & ... \\
2MASS $J$ & 12.20$\pm$0.02 & 16.24$\pm$0.10 &  11.96$\pm$0.02 & 15.29$\pm$0.06 & 14.15$\pm$0.03  & 15.53$\pm$0.06  & 10.98$\pm$0.02 & 15.39$\pm$0.05  \\
2MASS $H$ & 11.63$\pm$0.03 & 15.77$\pm$0.14 & 11.38$\pm$0.03 &14.57$\pm$0.07 & 13.56$\pm$0.04 & 14.85$\pm$0.07 & 10.41$\pm$0.02 & 14.64$\pm$0.05  \\
2MASS $K$ &  11.40$\pm$0.02 & 15.28$\pm$0.19 &  11.10$\pm$0.03 &13.95$\pm$0.07 & 13.21$\pm$0.04  & 14.59$\pm$0.08  & 10.16$\pm$0.02 & 14.16$\pm$0.07 \\
  & \multicolumn{2}{c}{-----------------------------------}   & \multicolumn{2}{c}{-----------------------------------}  &  \multicolumn{2}{c}{-----------------------------------}   & \multicolumn{2}{c}{-----------------------------------} \\
     Separation  &\multicolumn{2}{c}{30$^{\prime\prime}$.4} &\multicolumn{2}{c}{7$^{\prime\prime}$.6} & \multicolumn{2}{c}{10$^{\prime\prime}$.1}   & \multicolumn{2}{c}{46$^{\prime\prime}$.0}   \\
 Sep. (AU) & \multicolumn{2}{c}{2979$\pm$304}  &  \multicolumn{2}{c}{374$\pm$76}  &\multicolumn{2}{c}{756$\pm$200} &  \multicolumn{2}{c}{2418$\pm$664}\\
   SP$^{\mathrm{b}}$ & \multicolumn{2}{c}{3.49$\times10^{-3}$} & \multicolumn{2}{c}{2.23$\times10^{-4}$} &\multicolumn{2}{c}{6.57$\times10^{-4}$} & \multicolumn{2}{c}{8.75$\times10^{-4}$}   \\
\hline
\end{tabular}
%\end{minipage}
\begin{list}{}{}
%\item[$^{a}$] NLTT 7059.
\item[$^{a}$] Distances calculation assumes unresolved equal-mass binaries.
\item[$^{b}$] Statistical probability (SP) that the common proper motion could occur by random chance, see section \ref{s5.3} for detail of the method.
\item[$^{c}$] Characterized by \citet{haw} as an L3 dwarf.
\item[$^{d}$] Catalogued by \citet{wes} as L0, L0, M9 and M8 dwarf.
\item[$^{e}$] The distance given by \citet{lep} is 32.5$\pm$6.8.
\item[$^{f}$] The star is too bright for UKIDSS, thus $JHK$ magnitudes are not reliable.
\item[$^{g}$] Based on optical -- near-infrared colours.
\item[$^{h}$] We used more digital number in their name to distinguish the two components, SDSS 1304, SDSS 1631, SDSS 1638 will reference to secondary ultra-cool dwarfs in this paper.
%\item[$^{g}$] Photometric metallicity based on $V-K$ and $M_{K}$ \citep{joh09}.
\end{list}
%\end{center}
\label{t8bi}
\end{table*}

\begin{table}
\caption{Parameters of SDSS 0838: a candidate Praesepe member.}
\begin{center}
\begin{tabular}{|l|l|l|}
\hline\hline
Parameter & & Value \\
\hline
RA (J2000) & .................. & 08 38 23.69\\
Dec (J2000) & .................. & 16 47 02.0 \\
Sp. Type &  ..................  & L0.5 \\
Dist$^{a}$ (pc)         & .................. & 181$\pm62$  \\
Dist.b$^{a}$ (pc)       & .................. & 256$\pm62$  \\
 $\mu_{\rm RA}$     (mas$/$yr )        & .................. & $-$21$\pm$15 \\
 $\mu_{\rm Dec}$ (mas$/$yr )       & .................. & $-$19$\pm$15  \\
SDSS $r$                & .................. & 24.28$\pm$0.53 \\
SDSS $i$                & .................. & 22.32$\pm$0.21  \\
SDSS $z$                & .................. & 20.41$\pm$0.18 \\
UKIDSS $Y$              & ..................  & 19.24$\pm$0.08   \\
UKIDSS $J$              & .................. & 18.29$\pm$0.06   \\
UKIDSS $H$              & ..................  & 17.74$\pm$0.07   \\
UKIDSS $K$              &  .................. & 17.02$\pm$0.06   \\
\hline
\end{tabular}
%\begin{list}{}{}
%\item[$^{a}$]Based on relationship between M$_{i}$ and $i - J$ \citep{haw}.
%\item[$^{b}$]Based on relationship between M$_{i}$ and $i - z$ \citep{haw}.
%\end{list}
\end{center}
\label{t0838}
\end{table}

\begin{table}
 \centering
 %\begin{minipage}{140mm}
  \caption{Parameters of $\eta$ Cancri AB.}
  \begin{tabular}{lll}
  \hline\hline

Parameter & Adopted Value & Reference/Note\\
\hline
Separation (AU)      & 15019$^{+562}_{-522}$  & This paper \\
                 & (164\arcsec.16)      &  \\
Distance (pc)            & 91.5$^{+3.5}_{-3.2}$         & van Leeuwen (2007)  \\
SP & 0.05 & This paper \\
\hline
$\eta$ Cancri A \\
\hline
Sp. Type            & K3III & Perryman (1997) \\
RA (J2000)      & 08 32 42.50 & SDSS (2004) \\
Dec (J2000)     & 20 26 27.7 & SDSS (2004) \\
$V$             & 5.330$\pm$0.004 & Kharchenko (2000) \\
$B-V$           & 1.252$\pm$0.002 & Perryman (1997) \\
$V-I$           & 1.11$\pm$0.04 & Perryman (1997) \\
$J$     & 3.49$\pm$0.30 & 2MASS \\
$H$     & 2.86$\pm$0.21 & 2MASS  \\
$K$     & 2.70$\pm$0.32 & 2MASS  \\
Parallax (mas)          & 10.93$\pm$0.40 & van Leeuwen (2007) \\
 $\mu_{\rm RA}$     (mas$/$yr)     & $-$46.33$\pm$0.43 & van Leeuwen (2007) \\
 $\mu_{\rm Dec}$ (mas$/$yr)        & $-$44.31$\pm$0.24 & van Leeuwen (2007) \\
$U$ (km$/$s)                & $-$27.1  & Famaey (2005) \\
$V$ (km$/$s)                & $-$24.6  & Famaey (2005) \\
$W$ (km$/$s)            & $-$10.7  & Famaey (2005) \\
$T_{\rm eff}$ (K)       & 4446$^{a}$$\pm$80   &  Luck (2007)  \\
                && Hekker (2007) \\
                & & Allende Prieto (1999) \\
log ($L/$L$_{\odot}$)   & 1.95$\pm$0.06 &   Luck (2007) \\
Mass     (M$_{\odot}$)  & 1.4$^{b}$$\pm$0.2       & Luck (2007) \\
                && Allende Prieto (1999) \\
Radius (R$_{\odot}$)        & 16.0$^{c}$$^{+1.8}_{-1.6}$   & This paper  \\
$g$ (cm/s$^{2})$    & 2.24$\pm$0.09 dex & Luck (2007) \\
$[$Fe/H$]$      & 0.0$^{d}$$\pm$0.1 & This paper  \\
Age (Gyr)       & 2.2$-$6.1 & This paper \\
\hline
   \multicolumn{2}{l}{$\eta$ Cancri B  (SDSS 0832)}  \\
\hline
Sp. Type            & L3.5$\pm$1.5$^{e}$ & This paper \\
RA (J2000)      & 08 32 31.87  & SDSS (2004) \\
Dec (J2000)     & 20 27 00.0  & SDSS (2004) \\
Dist (pc)           & 121$\pm33$  & This paper \\
Dist.b  (pc)            & 171$\pm33$ & This paper \\
$\mu_{\rm RA}$     (mas$/$yr )        & $-$47.4$\pm$8.2 & This paper \\
 $\mu_{\rm Dec}$ (mas$/$yr )       & $-$38.9$\pm$8.2  & This paper \\
$r$         & 25.37$\pm$0.63 & SDSS \\
$i$             & 22.13$\pm$0.17 & SDSS \\
$z$             & 20.15$\pm$0.12 & SDSS\\
$Z$             & 19.69$\pm$0.09 & UKIDSS \\
$Y$             & 18.66$\pm$0.05 & UKIDSS  \\
$J$             & 17.74$\pm$0.04 & UKIDSS  \\
$H$             & 17.16$\pm$0.04 & UKIDSS  \\
$K$             & 16.52$\pm$0.04 & UKIDSS  \\
Mass (M$_{Jup}$)      & 63$-$82 & $^{e}$  \\
$T_{\rm eff}$ (K)       & 1800$\pm$150 & $^{e}$    \\
$g$ (cm/s$^{2})$    & 5.3$-$5.5 dex & $^{e}$ \\
\hline
\end{tabular}
\begin{list}{}{}
\item[$^{a}$] This is the average $T_{\rm eff}$ of Luck (2007),  Hekker (2007) and Allende Prieto (1999). $^{b}$ This is average mass of Luck (2007)  and Allende Prieto (1999). $^{c}$ Based on log ($L/$L$_{\odot}$) \citep{luc} and $T_{\rm eff}$ (4446 K). Note that the radius given in \citet{all} is 17.0$^{+2.5}_{-2.2}$ R$_{\odot}$ along with a lower $T_{\rm eff}$ (4365 K). $^{d}$ This is medium value of Luck (2007),  Hekker (2007) and Brown (1989). $^{e}$ Based on Lyon duty model \citep{cha00} and \citet{burr06,burr97} model, assume as a single object at the same distance with $\eta$ Cancri A.
\end{list}
%\end{minipage}
\label{tcnc}
\end{table}

\subsection{Common proper motion systems}
\label{s5.3}

We searched for common proper motion companions to our ultra-cool dwarfs using three methods that combine efficacy with efficiency. Our searches are not fully complete, but are productive and effective in the identification of companions.

For our sample of 271 objects with proper motions larger than 100 mas/yr, we searched for common proper motion companions in high proper motion catalogues \citep{lep,iva} out to 3\arcmin~separation, which corresponds to separation $<$9000 AU for closer objects (50 pc) and $<$27000 AU for more distant objects (150 pc), and avoids an excess of contamination through chance alignments (see Section 5.4). Four of the ultra-cool dwarfs with proper motions greater than 100 mas/yr were found to have common proper motion companions (allowing for astrometric uncertainties), and the distance constraints for each pair were consistent with shared distance. Three of these four objects have SDSS spectra (SDSS 0207, SDSS 0858 and SDSS 0956)  and one [SDSS J095324.54+052658.4 (SDSS 0953)] does not.

We also conducted a systematic searched for companions to our lower proper motion ($<$100mas/yr) objects. This search was conducted by visiual inspection of the region of sky around each of our lower proper motion ultra-cool dwarfs using the SDSS Navigate Tool. We inspected images covering separations out to 2\arcmin~on the sky, and looked for objects that could be mid K to mid M companions ($r-i\sim$0.5-2.0, $i-z\sim$0.2-1.0) with distances comparable to our sample ($\sim$30-150pc), and thus magnitude of $i\sim$9-17. We then measured the proper motions of all such possible companions (as described in Section 5) to test for companionship. In this way, another four objects were found to have nearby common proper motion companions with reasonable distance agreement, three of which have SDSS spectra (SDSS 1304, SDSS 1631, SDSS 1638) and one of which [SDSS 010153.11+152819.4 (SDSS 0101)] does not.

In addition, we performed a wider angle visual search for bright companions to the two faint low proper motion ultra-cool dwarfs towards Praesepe, out to 3\arcmin separation. This resulted in the identification of a bright giant star ($\eta$ Cancri) 164$^{\prime\prime}$ away from SDSS 0832. The proper motion of $\eta$ Cancri ( $\mu_{\rm RA}$ = $-$46.33$\pm$0.43 mas/yr,  $\mu_{\rm Dec}$ = $-$44.31$\pm$0.24 mas/yr) is taken from the updated $Hipparcos$ catalogue \citep{van} and is consistent with SDSS 0832 to within the astrometric uncertainties. Figure \ref{pm3} shows the proper motion of SDSS 0832 and $\eta$ Cancri (as well as SDSS 0838 and SDSS 1638). Figure \ref{eta} shows a UKIDSS \emph{JHK} false colour image of SDSS 0832 and $\eta$ Cancri.

We also obtained all available proper motions from the literature (VizieR/CDS database) for comparison (summarise in Table \ref{t8pm}), and note just one minor inconsistency. The literature proper motion of the companion (SDSS J163814.32+321133.5) to SDSS 1638 ( $\mu_{\rm RA}$ = $-$61$\pm$10,  $\mu_{\rm Dec}$ = 9$\pm$10) was seen to be somewhat smaller than our initial database-measured proper motion ( $\mu_{\rm RA}$ = $-$85$\pm$10,  $\mu_{\rm Dec}$ = 2$\pm$10). We therefore re-measured the proper motion of SDSS J163814.32+321133.5 and SDSS 1638 using a careful selection of fiducial reference stars and the {\scriptsize IRAF} routines {\scriptsize GEOMAP} and {\scriptsize GEOXYTRAN}. The results ( $\mu_{\rm RA}$ = $-$68$\pm$15,  $\mu_{\rm Dec}$ = $-$4$\pm$12 for  SDSS J163814.32+321133.5;  $\mu_{\rm RA}$ = $-$82$\pm$15,  $\mu_{\rm Dec}$ = 9$\pm$12 for SDSS 1638) show a good agreement with literature values, and we conclude that in this case the proper motion measured direct from the database coordinate/epoch information was not optimal.

\begin{figure*} %  figure placement: here, top, bottom, or page
   \centering
   \includegraphics[width=500pt, angle=0]{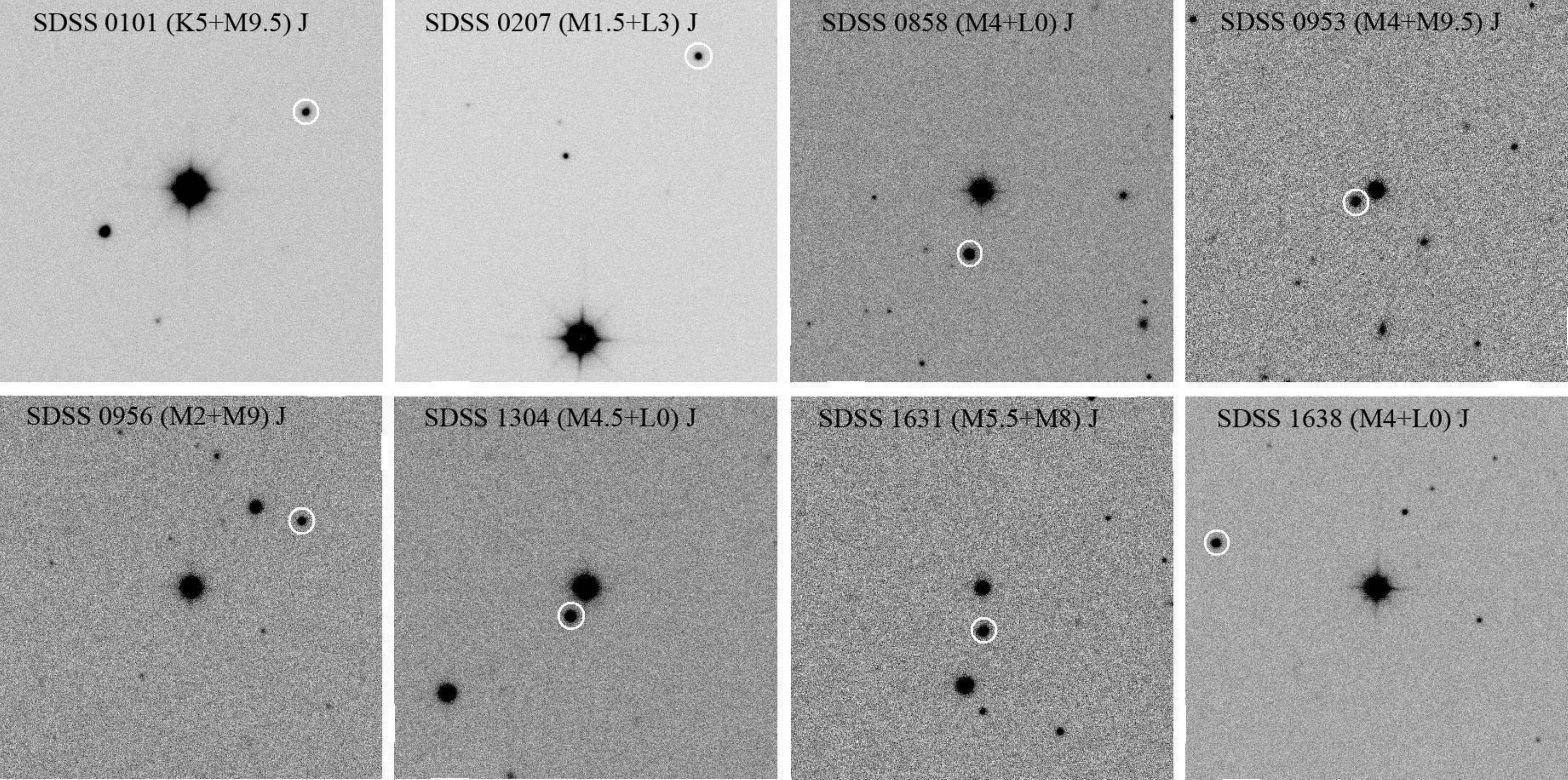}
   \caption{UKIDSS \emph {J}-band images of eight ultra-cool dwarf binary systems. All the images have a size of 1.5 arcmin with north up and east left. The primary stars are generally in the centre of imges, except the second one (SDSS 0207) in which the primary star is in the bottom of the field, the ultra-cool dwarfs are in \emph{white circles}.}
   \label{bi8}
\end{figure*}

Overall, we thus found nine star + ultra-cool dwarf common proper motion pairs. Proper motions of seven of these are shown in Figure \ref{pm7}, and the proper motions of the other two (SDSS 0832 and SDSS 1638) are shown in Figure \ref{pm3} for clarity. Table \ref{t8bi} shows the parameters of the eight dwarf star + ultra-cool dwarfs common proper motion pairs, and Figure \ref{bi8} shows UKIDSS $J-$band images of these systems. The $\eta$ Cancri system is discussed further in Section 6.
 
\begin{figure} %  figure placement: here, top, bottom, or page
   \centering
   \includegraphics[width=\columnwidth]{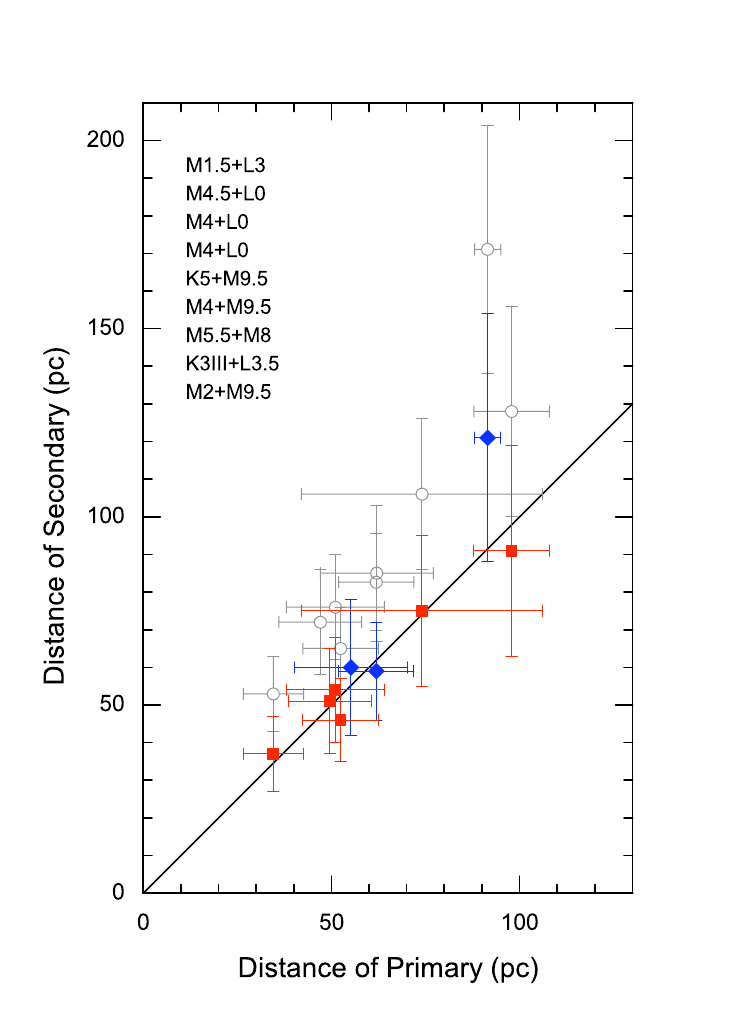}
   \caption{Distances of ultra-cool dwarfs (from left to right: SDSS 0207, 1304, 1638,  0858, 0953, 0101, 1631,  0832, 0956) and their primary stars. Spectral types of combinations are listed in according distance of primaries in the top left of the plot. The \emph{red squares} and \emph{blue diamonds} indicate the objects with and without SDSS spectra respectively. The \emph{gray open circles} show the distances of ultra-cool dwarfs assume as unresolved equal-mass binaries.}
\label{dd9}
\end{figure}

\subsection{Chance alignments?}
\label{s5.4}

To determine if the new binary systems are genuine, we carried out a statistical analysis of the probability that the pairs could be line-of-sight associations with photometry and proper motion consistent with binarity by random chance (using a similar method to that used by \citet{day}). As an example we discuss the case of $\eta$ Cancri and SDSS 0832. To assess the probability of common proper motion we searched the SuperCOSMOS Science Archive \citep{ham01} in a square degree of sky around the position of the L dwarf, and made a colour-magnitude selection (B versus B$-$R) consistent with a dwarf sequence in the $\sim$90-160 pc distance range. We found that seven out of 337 objects had common proper motion to within 2 sigma uncertainty, suggesting a probability of 2 per cent that such common proper motion could occur by random chance. Separations out to 164\arcsec\ from $\eta$ Cancri and a distance range of 90$-$160 pc correspond to a space volume of 2.86 pc$^{3}$. Assuming a space density of 0.0106 pc$^{-3}$ for $\sim$G-K stars  \citep[M$_V$=5.5-8.5;][]{kro95}, we thus estimate that we would expect 4.9$\times10^{-4}$ G or K stars to masquerade as common proper motion companions in the volume considered around the L dwarf SDSS 0832.

Our full ultra-cool dwarf sample contains 268 objects with proper motion $>$ 100 mas/yr, 283 with proper motion = 50-100 mas/yr, and 258 with proper motion $<$ 50 mas/yr. $\eta$ Cancri resides in our lowest proper motion group, and companions to ultra-cool dwarfs in this group were searched for (in all but 2 cases) out to separations of 2\arcmin. Accounting for the typical volume searched around each ultra-cool dwarf, and the total number of ultra-cool dwarfs in this group, we estimate that we might have expected to find $\sim$0.05 apparent $\eta$ Cancri-like common proper motion companions by random chance. We performed similar calculations for our other binary systems, using a K-M space density of 0.04 pc$^{-3}$ \citep{kro95} and accounting for the significantly smaller separations and generally larger proper motions (see Table \ref{t8bi}). We note that the expected number of random chance alignments for these other systems (0.00002$-$0.0048) are much less than for the $\eta$ Cancri system. This analysis shows that overall we would not expect any random chance common proper motion companions to our full ultra-cool dwarf sample, and that our nine identified systems should be genuine binaries. However, we note that for the wide separation and lower proper motion of $\eta$ Cancri B, the possibility of spurious companionship is larger (although still very small) than for the other binaries.

\subsection{New binaries}
\label{s5.5}

Figure \ref{dd9} shows the distance constraints of the nine binary ultra-cool dwarfs plotted against the distance constraints for their respective primary stars. Distances of the seven brighter primary dwarfs are based on the relationship between the \emph{V}$-$\emph{J} colour index and absolute magnitude $M_{V}$ \citep{lea}. However, SDSS J163126.17+294847.1 (M5.5V) is too faint and red to have a $V$-band magnitude available, so we constrained its distance using the same method that was used for the ultra-cool dwarfs (see Section 4). The distance of $\eta$ Cancri is from $Hipparcos$ \citep{van}. This plot allows us to place constraints on possible unresolved binarity amongst our ultra-cool dwarf companions. It can be seen that six ultra-cool dwarfs could be equal-mass unresolved binaries by the distance constraints. We can't, however, rule out the possibility of non-equal-mass unresolved binarity from this plot. If SDSS 0832 for example was an equal-mass binary, it would be too distant to be associated with $\eta$ Cancri. However, a non-equal-mass binary (e.g. an unresolved L + T binary) would affect spectral type and distance estimates as well as absolute magnitude, and this is not ruled out by Figure \ref{dd9}. This will be discussed further in Section \ref{s6.3}.

\begin{figure} %  figure placement: here, top, bottom, or page
   \centering
   \includegraphics[width=\columnwidth]{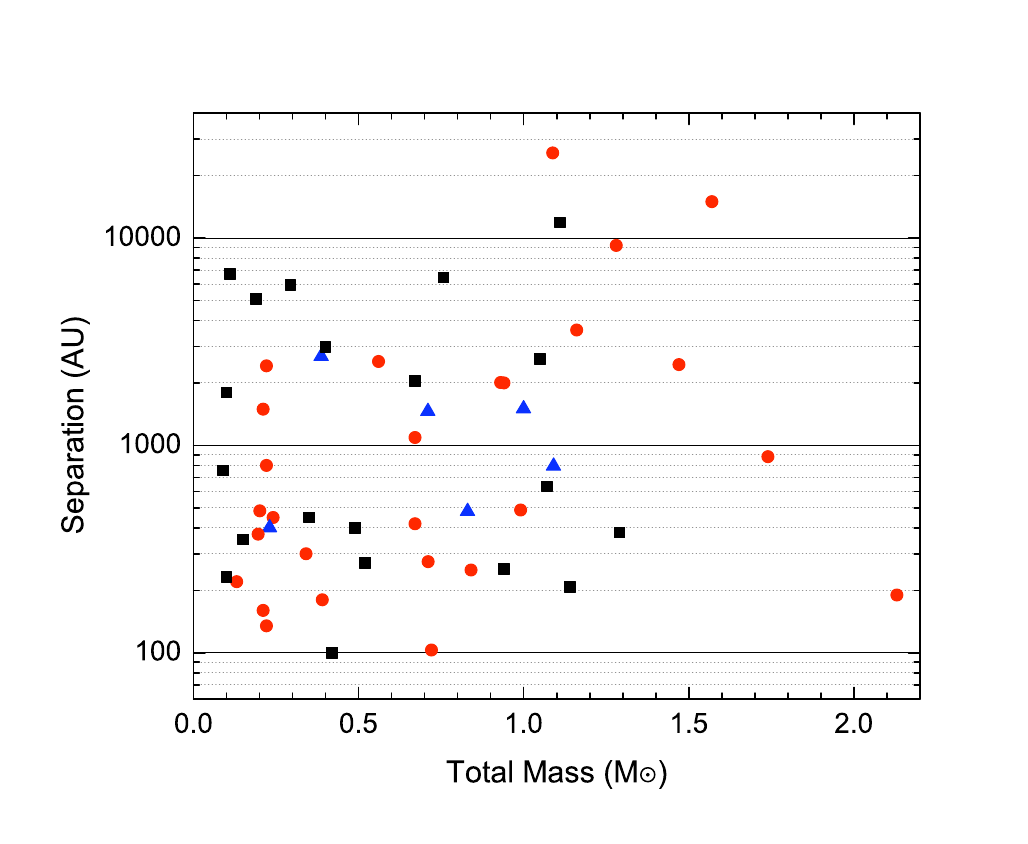}
   \caption{Total mass versus separation for binaries have projected separation larger 100 AU. \emph{Black squares}, \emph{red circles} and \emph{blue triangles} are binaries with late M, L and T secondaries.} 
\label{tms}
\end{figure}

Figure \ref{tms} shows our new ultra-cool dwarf + star binaries in the context of the broader population of such wide systems \citep[we plot known binaries with a$>$100AU; see][]{fah10,art09,bur,rad09,rad08,pha08,cab07,cru,luh,mug07,met06,rei06,bil05, cha05,neu05,neu04,sei05a,sei05b,gol04,sch04,sch03,bou03,giz01,giz00,wil01,bur00,low00,whi99,reb98}. It is clear from this plot that the projected separation range $<$7000AU is well populated for binaries with a total mass $<$1M$_{\odot}$, with wider ultra-cool dwarf companions to such stars being rare. However, there is some indication that higher mass primaries are able to host wider ultra-cool dwarf companions, with the $\eta$ Cancri system contributing to this trend.

\section{Eta Cancri AB}

With a separation of 15019 AU, $\eta$ Cancri AB is an extremely wide binary, however, as Section 5.4 explains, it is a statistically solid association, with the two components presumably having a common composition and age. Only five brown dwarf mass objects were previously known around (four) giant stars, all detected by the radial velocity method \citep{hat,lov,liu,nie}. These objects all have $m$ sin $i$ in the range 10$-$20 M$_{Jup}$ and are within three AU of their primaries. They are thus unobservable with current technology, and their mass range overlaps that of the planetary regime. As such, their atmospheres may not share the same composition as the primary star, and without direct observation they are not useful benchmark objects for study.

\begin{figure*} %  figure placement: here, top, bottom, or page
   \centering
   \includegraphics[height=288pt, angle=0]{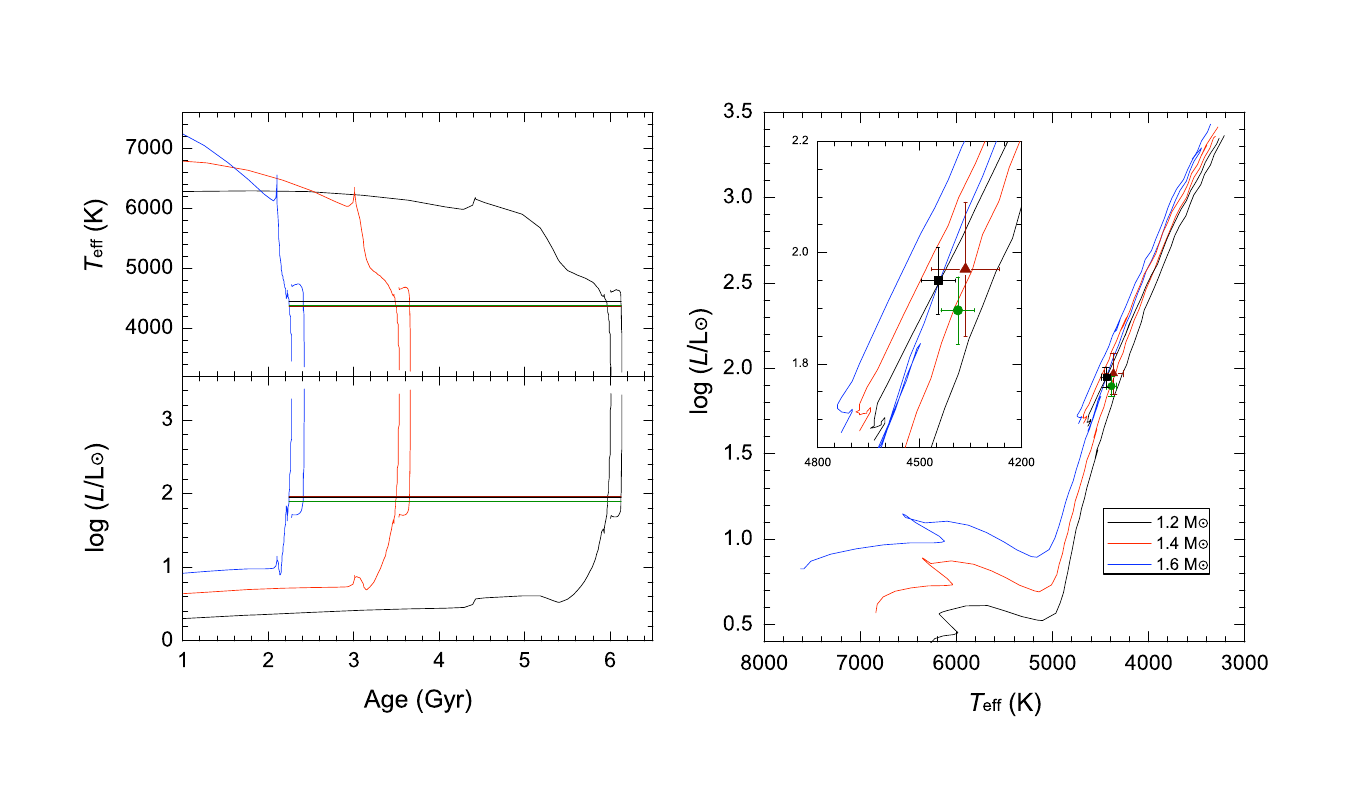}
   \caption{$T_{\rm eff}$ and \emph{L} evolutionary tracks \citep{gir} of solar metallicity stars with different masses (1.2M$_{\odot}$, \emph{black lines}; 1.4M$_{\odot}$, \emph{red lines}; 1.6M$_{\odot}$, \emph{blue lines}). The \emph{filled black square} (4446, 1.95) in the right panel represents the average $T_{\rm eff}$ from the literature and \emph{L}  from \citet{luc}. The \emph{filled green circle} (4388, 1.90) represents the values from PARAM \citep{sil}. The \emph{filled brown triangle} (4365, 1.97) represents the values from \citet{all}. \emph{Horizontal lines} in left hand plots indicated these \emph{L} and $T_{\rm eff}$ values in the same colours with symbols in right hand plots.}
   \label{alt}
\end{figure*}

\begin{figure*} %  figure placement: here, top, bottom, or page
   \centering
   \includegraphics[width=320pt]{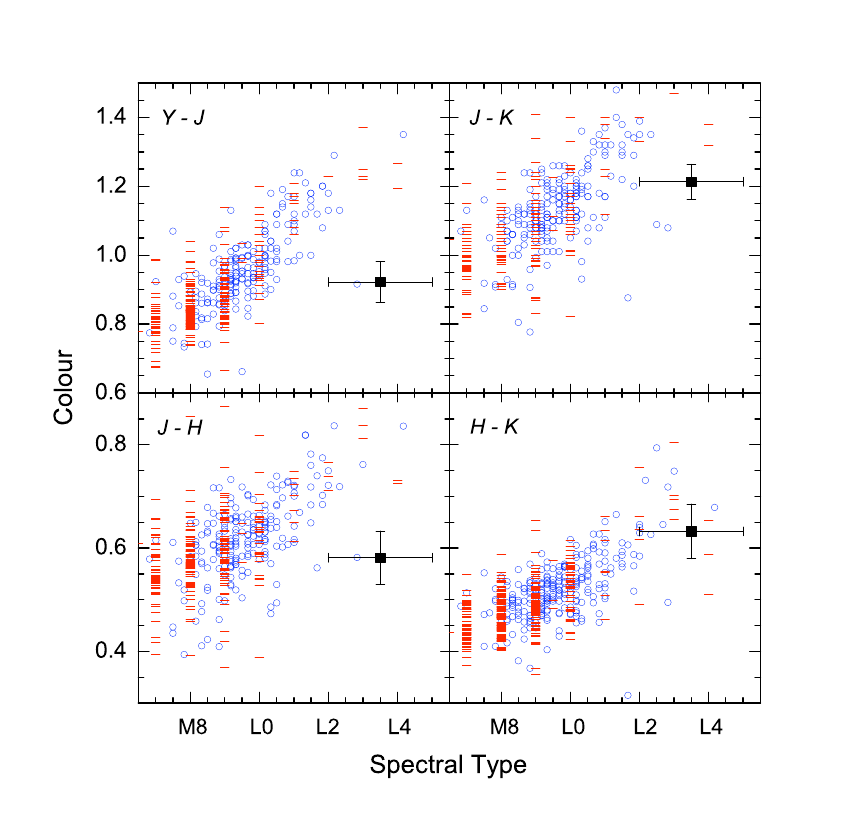}
   \caption{UKIDSS \emph{YJHK} colours of $\eta$ Cancri B (\emph{black squares}) and ultra-cool dwarfs with (\emph{red bars}) and without (\emph{blue circles}) spectra, as a function of spectral type.}
\label{col}
\end{figure*}

\begin{figure*} %  figure placement: here, top, bottom, or page
   \centering
   \includegraphics[width=450pt]{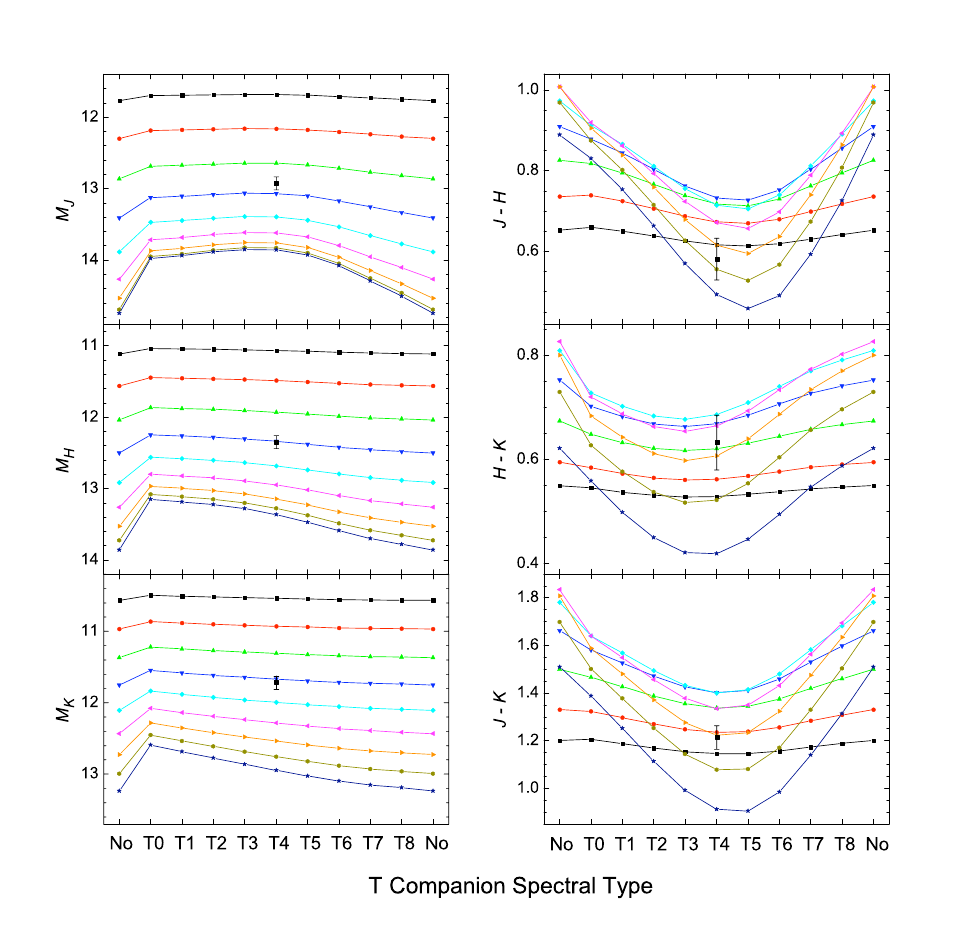}
   \caption{\emph{JHK} absolute magnitudes and colours of unresolved L+T binaries. Different \emph{lines} represent absolute magnitudes and colours of L dwarf primaries (L1$-$L9, from top to bottom in left panel) without or with different T companions (T0$-$T8). If $\eta$ Cancri B (\emph{black square}) is actually an unresolved binary, then the constituents would likely be made up of an L4 and a T4 dwarf.}
\label{bmc}
\end{figure*}

\subsection{The metallicity of $\eta$ Cancri A}

$\eta$ Cancri A is a bright K3 giant with a \emph{V} band magnitude of 5.33, and it's metallicity  has been measured several times by various groups; [Fe/H] = 0.19 by \citet{bro}, [Fe/H]=0.07$\pm$0.21, [Fe/H] = 0.08, [{Fe~\sc{i}}/H] = 0.02, [{Fe~\sc{ii}}/H] = 0.10 measured with different methods by \citet{luc} and $A(\mathrm{Fe})=7.64$ by \citet{hek}. These results are consistent with a metallicity of [Fe/H]=0.10$\pm$0.1 for $\eta$ Cancri A. Given the importance of being able to infer a metallicity for the L dwarf companion, we have re-addressed the metallicity of this system with new data and analysis.

We obtained a high resolution echelle spectrum of $\eta$ Cancri A during an observing run from 9 to 11 February 2009, using the FIber fed Echelle Spectrograph (FIES) at the 2.56~m Nordic Optical Telescope located at the Observatorio del Roque de los Muchachos (La Palma, Spain). We used FIES with a 2048$\times$2048 13.5$\mu$m pixel EEV42-40  CCD detector. The wavelength range covered is from 3620 to 7360 {\AA} in 80 orders. The reciprocal dispersion ranges from 0.02 to 0.04 {\AA}/pixel and the spectral resolution (FWHM) is from 0.06 to 0.10 {\AA}. We extracted the spectrum using standard reduction procedures in IRAF, including bias subtraction, flat-field division and optimal extraction of the spectra. We obtained the wavelength calibration by taking a spectrum of a Th-Ar lamp. Finally, we normalized the spectrum by a polynomial fit to the observed continuum.

Stellar iron abundances [$A(\mathrm{Fe})=\log{N(\mathrm{Fe})/N(\mathrm{H})}+12$] were estimated from the equivalent width measurements of a set of thirteen {Fe~\sc{i}} and two  {Fe~\sc{ii}} lines. The line list was provided by Jorge Mel{\'e}ndez (private communication) who has tested their use for F, G and K giant stars. Measurements were compared with equivalent widths computed under the assumption of local thermodynamical equilibrium (LTE) and spherical symmetry, which for stars of low surface gravity should be more appropriate than a plane-parallel description. Calculations were made using the Uppsala suite of codes and the new MARCS model atmospheres \citep{gus08}. Stellar atmospheres are characterised by four fundamental parameters: $T_\mathrm{eff}$, $\log{g}$, metallicity ([Fe/H] = $A(\mathrm{Fe})-A(\mathrm{Fe})_{\sun}$) and microturbulence ($\xi_t$). For the values of interest here, a grid of model atmospheres of one solar mass and standard composition was linearly interpolated. The $T_\mathrm{eff}$ of $\eta$ Cancri A was estimated from Johnson photometry in the blue ($V=5.343$ and $B=6.608$) and 2MASS in the red ($J=3.489$, $H=2.861$ and $K=2.699$), using the \citet{ram05} $T_\mathrm{eff}$-vs-$(B-V)$, $(V-H)$, $(V-J)$ and $(V-K)$ calibrations. Estimates coming from the blue and the red were somewhat different, 4329~K versus 4520~K, so an average value of 4472~K with an assigned 107~K uncertainty in our computations. The interstellar extinction of this nearby star estimated from the models in \citet{hak97} is very low $E(B-V)=0.01\pm0.05$ and it cannot explain the temperature discrepancy between these two estimates. Also we tried the new $T_\mathrm{eff}$-vs-colour calibrations in \citet{gon09} but the difference is even larger ($T_\mathrm{eff}~4531\pm137$) with the blue estimate. Another concern is the precision of the surface gravity determination which in turns compromises the accuracy of the metallicity determinations. Our $\log{g}=2.30$ estimate is based on the Hipparcos parallax of the star ($\pi=10.93\pm0.43$, \citealt{van}), and on a $M\sim1.60~M_{\odot}$ stellar mass and a $BC=-0.558$ bolometric correction for $V$. These two last values are estimated from isochrones and were taken from \citet{luc} after checking they were appropriate for our stellar temperature and metallicity values (for details on the accuracy of these parameter determinations see \citet{luc}). Another way of estimating the surface gravity is the Fe ionization equilibrium, however, this method has the disavantage of being sensitive to departures from LTE. In our case, our best estimate of $\log{g}$ (2.30) produces an abundance difference between {Fe~\sc{i}} and {Fe~\sc{ii}} of ${\Delta}A(\mathrm{Fe})=0.15$~dex. The microturbulence value was changed until the trend in $A$(Fe)-vs-Fe\texttt{I} equivalent widths was minimised, which happened at $\xi_t=1.62~\mathrm{km/s}$.

The method employed here requires the knowledge a priori of the metallicity so that two iterations were necessary to get our best iron abundance value of 7.52~dex (an average value over the fifteen Fe lines). Results based on the {Fe~\sc{ii}} lines are sensitive to the uncertainties of the stellar parameters: abundance changes of $-0.12$~dex for $\Delta T_\mathrm{eff}\sim100$~K, $0.08$~dex for $\Delta\log{g}\sim0.15$ (mass, parallax), $0.05$~dex for $\Delta$[Fe/H]$\sim$0.10 dex. However, the average abundance based on the neutral lines is much less sensitive (abundance changes $\sim$ 0.02-0.03~dex). Changes in the microturbulence and the uncertainties in the measured equivalent widths have a more significant effect, 0.05~dex for 0.1 $\mathrm{km/s}$ microturbulence changes, and a similar effect for a typical ${\Delta}W_{\lambda}=3$ m{\AA} uncertainty. We have adopted our total error budget to be $\pm$0.10~dex, the standard deviation associated with the fifteen lines used, which will partially account for the sensitivity mentioned before. We also note that our abundances may suffer from other systemactic errors, e.g. $T_\mathrm{eff}$ zero points, isochrones uncertainties, departures from LTE. Our final metallicity value depends on the solar iron abundance that we assume; if the Sun was analysed in a similar way to $\eta$ Cancri A (i.e. 1D-LTE but using the appropriate line list e.g. \citealt{gar06}), then the metallicity (relative to solar) of $\eta$ Cancri A would be $-0.02$ dex [$A(\mathrm{Fe})_{\odot}=7.54$]. Solar abundances based on 3D-LTE analysis are much lower, 7.45 \citet{gre07}, which leads to higher metallicity value for $\eta$ Cancri A of 0.07 dex. On balance we present a metallicity constraint (in Table \ref{tcnc}) for $\eta$ Cancri A of 0.0$\pm$0.1.

\subsection{The age of $\eta$ Cancri A}

The $T_{\rm eff}$ measurements of $\eta$ Cancri A measured here and by \citet{all, luc, hek} are all consistent with $T_{\rm eff}$ = 4450$\pm$100 K. The radius of $\eta$ Cancri A based on this temperature and log ($L/$L$_{\odot}$) = 1.95$\pm$0.06 \citep{luc} is 16.0$^{+1.8}_{-1.6}$ R$_{\odot}$. \citet{luc} estimated a mass for $\eta$ Cancri A of 1.6$\pm$0.5 M$_{\odot}$ (typical uncertainty $\sim$35\%), and \citet{all} estimates a mass of 1.36$\pm$0.67 M$_{\odot}$. To provide additional refinement, we calculated the parameters of $\eta$ Cancri A with  PARAM \citep{sil}. PARAM is a web interface for the Bayesian estimation of stellar parameters. Table \ref{tparam} shows our input and output values from PARAM. The $[$Fe/H$]$,  $V$ magnitude, parallax, and $B-V$ colour of $\eta$ Cancri A are known, so we experimented with different $T_{\rm eff}$ input values until the model colour output was $B-V = 1.252$ \citep{per}. The mass, gravity and radius output from PARAM are all consistent with other estimates, but provide a somewhat improved level of accuracy, and on balance we assumed a mass of 1.40$\pm$0.2 M$_{\odot}$ for $\eta$ Cancri A.

\begin{table}
 \centering
\caption{Input and output values from PARAM for $\eta$ Cancri.}
%\begin{center}
\begin{tabular}{|l|l|l|l|l |}
\hline\hline
\multicolumn{2}{c}{Input value}  & & \multicolumn{2}{c}{Output value} \\
\hline
$T_{\rm eff}$   & 4388$^{a}$$\pm$80 K && Age    & 3.61$\pm$1.86 Gyr \\
$[$Fe/H$]$$^{b}$    & 0.025$\pm$0.100 && Mass & 1.36$\pm$0.23 M$_{\odot}$\\
$V$$^{b}$ & 5.330$\pm$0.004 && $g$  & 2.16$\pm$0.11 dex (cm/s$^{2}$) \\
Parallax$^{b}$ & 10.93$\pm$0.40mas && $R$   & 15.4$\pm$1.1 R$_{\odot}$ \\
\hline
&&& $B-V$           & 1.252$^{c}$$\pm$0.038 \\
\hline
\end{tabular}
\begin{list}{}{}
\item Note: PARAM is a web interface for the Bayesian estimation of stellar parameters, http://stev.oapd.inaf.it/cgi-bin/param \citep{sil}.
\item[$^{a}$] This is an estimated value to get the right $B-V$ value.
\item[$^{b}$] Values are from literatures.
\item[$^{c}$] This value equal  to observed value.
\end{list}
%\end{center}
\label{tparam}
\end{table}

We have used these physical properties to constrain the age of $\eta$ Cancri A by comparison to evolutionary tracks. Figure \ref{alt} shows the  $T_{\rm eff}$ and $L$ evolutionary tracks \citep{gir} as a function of time and on a Hertzsprung$-$Russell (HR) diagram for solar metallicity stars with different masses (1.2M$_{\odot}$, 1.4M$_{\odot}$, and 1.6M$_{\odot}$). $T_{\rm eff}$ and log ($L/$L$_{\odot}$) measurements from the literature are plotted in the HR diagram with different symbols (see caption). In the left-hand plots we indicate with open symbols where the measured values of luminosity and  $T_{\rm eff}$ intersect with the evolutionary model tracks. In this way the \citep{gir} evolutionary models lead to an age constraint of 2.2$-$6.1 Gyrs.

We have also used the \citet{hur} stellar evolution sequence with initial masses 1.2$-$1.6 M$_{\odot}$ ($\Delta$M $=$ 0.05 M$_{\odot}$) and metallicity [Fe/H] $=$ 0.0 and 0.10. Using the same approach as in Figure \ref{alt}, these models give an age range of 2.4$-$6.6 Gyr for $\eta$ Cancri A which is very similar to the result from the \citet{gir} model. In addition we note that the age estimate from PARAM is also consistent with these values estimated directly from evolutionary tracks. A summary of the properties of $\eta$ Cancri A is given in Table \ref{tcnc}.

\subsection{The properties of $\eta$ Cancri B}
 \label{s6.3}
{\bf As a single ultra-cool dwarf: }
We can assume that the distance of $\eta$ Cancri B is essentially the same as that of $\eta$ Cancri A. Although since this is a very wide binary system we have increased the uncertainty of the companion distance slightly to account for $\eta$ Cancri B potentially being $\sim$ 0.5 pc closer or further than the primary along the line of sight. This allows us to place resonably tight constraints on the absolute brightness of this ultra-cool dwarf. $\eta$ Cancri B must have $M_{J}$ = 12.93$\pm$0.09, $M_{H}$ = 12.35$\pm$0.09 and $M_{K}$ = 11.72$\pm$0.09. Using the $M_{JHK}-$ spectral type relations of \citet{lium} we thus estimate that its spectral type must be L3-4 if it is a single L dwarf. Combining these magnitudes and an age (inferred from $\eta$ Cancri A), the evolutionary model of \citet{cha00} predicts that mass = 63$-$74 M$_{Jup}$, $T_{\rm eff} =1805\pm$115 K and log ($g/$cms$^{-2})=5.31$$-$5.36. Using an alternative model set (Burrows et al. 2006) yields a slightly higher mass or 75$-$82M$_{Jup}$, and corresponding $T_{\rm eff} =$1800$\pm$150 K, log ($g/$cms$^{-2})=5.38$$-$5.48.

In Figure \ref{col} we compare the colour of $\eta$ Cancri B to our large population of SDSS ultra-cool dwarfs, and it can be seen that its colours appear somewhat discrepant from other early L dwarfs. While it has similar H-K to the bulk population, its $J-K$ colour is only similar to the bluer L dwarfs and its $Y-J$ and $J-H$ colours are significantly bluer than typical ultra-cool dwarfs of similar spectral type.  Theoretical model colour trends suggest that L dwarfs with low metallicity or higher gravity could have bluer near-infrared colours \citep{burr06}. However, the constraints we have placed on the properties of this benchmark object are not consistent with this interpretation, and the colours of $\eta$ Cancri B are either at odds with the model trends, or are indicative of unresolved multiplicity.

{\bf As an unresolved binary ultra-cool dwarf:}
We investigated the possibility that the colours of $\eta$ Cancri B may result from unresolved binarity. T dwarfs have slightly bluer $Y-J$ and significantly bluer $J-H$ and $J-K$ than L dwarfs \citep{kna04,pinf}. Combining the light of an L and T dwarf could thus result in these colours being bluer than those of a single L dwarf. We experimented by combining the photometric brightnesses of a variety of L and T dwarf combinations, using the spectral type$-$M$_{JHK}$ polynomial relations from \citet{lium} to provide absolute magnitudes for flux combination. Figure \ref{bmc} shows the \emph{JHK} absolute magnitudes and colours of the range of unresolved L-T binaries that we considered. The location of $\eta$ Cancri B is indicated with it appropriate uncertainties. Considering the full range of colour and magnitude criteria, as well as the intrinsic scatter in the \citet{lium} polynomial fits, we deduce that a combination of an L4 and T4 dwarf at the distance of $\eta$ Cancri A would have colours and magnitudes reasonably consistent with those of $\eta$ Cancri B. In this case, the physical parametres of these two unresolved objects (estimated from the same evolutionary models as before) are given in Table \ref{tlt}.

\begin{table}
 \centering
\caption{Parameters of an L4 and a T4 dwarf from the Lyon group and Burrows group models.}
%\begin{center}
\begin{tabular}{l l l l}
\hline\hline
 Spectral Type &L4$^{a}$ & T4$^{b}$ \\
\hline
$M_{J}$$^{c}$ & 13.41$\pm$0.37 &  14.49$\pm$0.37 \\
 $M_{H}$$^{c}$ & 12.50$\pm$0.30 &  14.46$\pm$0.30 \\
$M_{K}$$^{c}$ & 11.75$\pm$0.34 &  14.52$\pm$0.34 \\
Mass$^{d}$ (M$_{Jup}$) &59$-$80  &40$-$70 \\
$T_{\rm eff}$$^{d}$ (K)  & 1675$\pm$145 &  1140$\pm$120 \\
 $g$$^{d}$ (cm/s$^{2})$ &   5.3$-$5.5 dex & 5.1$-$5.4 dex \\
\hline
\end{tabular}
\begin{list}{}{}
%\item Note: These parameters are for an L4 dwarf and a T4 dwarf at age of 2.2$-$3.8 Gyrs.
\item[$^{a}$] Based on dusty model.
\item[$^{b}$] Based on dust-free model.
\item[$^{c}$] Based on the relationship between spectral types and absolute magnitudes ($J$- and $K$-band) of \citet{lium}.
\item[$^{d}$] These parameters are for an L4 dwarf and a T4 dwarf at age of 2.2$-$6.1 Gyrs.
\end{list}
%\end{center}
\label{tlt}
\end{table}

\section{Summary and future work}

We have identified 34 new L dwarfs and nine ultra-cool dwarf + star common PM binary systems (TYC 1189-1216-1 AB, G 3-40 AB, LP 312-49 AB, LP 548-50 AB, LP 609-3 AB, SDSS J130432.93+090713.7 AB, SDSS J163126.17+294847.1 AB and  SDSS J163814.32+321133.5 AB, $\eta$ Cancri AB). Further observations (parallax, activity and rotation rate) of the K-M dwarf primaries in these systems will provide age (e.g. Faherty et al. 2009) and metallicity (e.g. Johnson \& Apps 2009) constraints. $\eta$ Cancri AB is a K3III + early L system with a projected separation of 15019 AU. $\eta$ Cancri A has a well constrained parallax distance, and we use new observations and the latest theory to estimate an age of 2.2$-$6.1 Gyr and a composition near solar for the system. We thus estimate that $\eta$ Cancri B (as a single object) has mass = 63$-$82M$_{Jup}$, $T_{\rm eff}=1800\pm$150 K, and log ($g/$cms$^{-2})=5.3$$-$5.5, based on a range of evolutionary models.

It is unclear at this stage if $\eta$ Cancri B is a single L dwarf with unusual colour or an unresolved L+T binary whose components have fairly typical colours when compared to the bulk population. Quality optical$-$infrared spectroscopy would address this issue in more detail, and establish if the binary explanation remains possible. In ddition adaptive optics observations could resolve the pair if their separation is $>$ 0.05$-$0.1 arcsec \citep[e.g. LHS4009;][]{mon06} corresponding to a separation $>$ 5$-$10 AU. If the ultra-cool dwarf is shown to be single then this would provide a surprising result when compared to the models. We could also establish $T_{\rm eff}$ and log $g$ more robustly by obtaining full optical to infrared measurements to constrain the bolometric flux and luminosity (e.g. Burningham et al. 2009). If $\eta$ Cancri B is a closely separated ultra-cool dwarf binary system, it could be possible to derive dynamical masses for the components over a time-scale of several years, and thus improve the benchmark quality of this object still further through more direct constraints on its mass and log $g$. A more detailed examination of the spectral morphology and multiplicity of this benchmark ultracool dwarf will in any event provide rigorous tests for solar metallicity dusty ultracool atmosphere models.

\section*{Acknowledgments}

Funding for the SDSS and SDSS-II has been provided by the Alfred P.
Sloan Foundation, the Participating Institutions, the National
Science Foundation, the U.S. Department of Energy, the National
Aeronautics and Space Administration, the Japanese Monbukagakusho,
the Max Planck Society, and the Higher Education Funding Council for
England. The SDSS Web Site is http://www.sdss.org/.
The SDSS is managed by the Astrophysical Research Consortium for the
Participating Institutions. The Participating Institutions are the
American Museum of Natural History, Astrophysical Institute Potsdam,
University of Basel, University of Cambridge, Case Western Reserve
University, University of Chicago, Drexel University, Fermilab, the
Institute for Advanced Study, the Japan Participation Group, Johns
Hopkins University, the Joint Institute for Nuclear Astrophysics,
the Kavli Institute for Particle Astrophysics and Cosmology, the
Korean Scientist Group, the Chinese Academy of Sciences (LAMOST),
Los Alamos National Laboratory, the Max-Planck-Institute for
Astronomy (MPIA), the Max-Planck-Institute for Astrophysics (MPA),
New Mexico State University, Ohio State University, University of
Pittsburgh, University of Portsmouth, Princeton University, the
United States Naval Observatory, and the University of Washington.

The UKIDSS project is defined in \citet{law}. UKIDSS uses the UKIRT Wide Field Camera (WFCAM; Casali et al. 2007) and a photometric system described in \citet{hew}. The pipeline processing and science archive are described in Irwin et al. (in preparation) and \citet{ham}. We have used data from the fourth data release. This publication makes use of data products from the Two Micron All Sky Survey. This research has made use of data obtained from the SuperCOSMOS Science Archive, prepared and hosted by the Wide Field Astronomy Unit, Institute for Astronomy, University of Edinburgh, which is funded by the UK Science and Technology Facilities Council.

Based on observations obtained at the Gemini Observatory, which is operated by the Association of Universities for Research in Astronomy, Inc., under a cooperative agreement with the NSF on behalf of the Gemini partnership: the National Science Foundation (United States), the Science and Technology Facilities Council (United Kingdom), the National Research Council (Canada), CONICYT (Chile), the Australian Research Council (Australia), MinistŽrio da Cincia e Tecnologia (Brazil) and Ministerio de Ciencia, Tecnolog'a e Innovaci—n Productiva  (Argentina). Based on observations carried out with the ESO New Technology Telescope (NTT). This paper has used observations made at the Nordic Optical Telescope (NOT), at the Spanish Observatory del Roque de los Muchachos of the Instituto de Astrof\'{i}sica de Canarias.

This research has made use of the VizieR catalogue access tool, CDS, Strasbourg, France.
Research has benefitted from the M, L, and T dwarf compendium housed
at DwarfArchives.org and maintained by Chris Gelino, Davy
Kirkpatrick, and Adam Burgasser. IRAF is distributed by the National Optical Observatory,
which is operated by the Association of Universities for Research in
Astronomy, Inc., under contract with the National Science Foundation. This research has benefitted from the SpeX Prism Spectral Libraries, maintained by Adam Burgasser at http://www.browndwarfs.org/spexprism. ZengHua Zhang is supported by the University of Hertfordshire Research Studentship. Zhanwen Han is supported by NSFC (grant No. 10821061). CGT is supported by ARC grant DP774000.

%\bibitem[\protect\citeauthoryear{et al.}{}]{}

\bsp

\label{lastpage}


\begin{thebibliography}{99}
\bibitem[\protect\citeauthoryear{Abazajian et al.}{2009}]{aba} Abazajian K.N. et al., 2009, ApJS, 182, 543
\bibitem[\protect\citeauthoryear{Adelman-McCarthy et al.}{2008}]{ade} Adelman-McCarthy J.K. et al., 2008, ApJS, 175, 297
\bibitem[\protect\citeauthoryear{Allende Prieto et al.}{1999}]{all} Allende Prieto C., Lambert D.L., 1999, A\&A, 352, 555
\bibitem[\protect\citeauthoryear{Artigau et al.}{2009}]{art09} Artigau \'{E}., Lafreni\`{e}re D., Lo\"{i}c A., Doyon R., 2009, ApJ, 692, 149 
\bibitem[\protect\citeauthoryear{Bihain et al.}{2006}]{bih} Bihain G., Rebolo R., B\'{e}jar V.J.S. et al., 2006, A\&A, 458, 805
\bibitem[\protect\citeauthoryear{Bill\`{e}res et al.}{2005}]{bil05} Bill\`{e}res M., Delfosse X., Beuzit J.-L. et al., 2005, A\&A, 440, L55
\bibitem[\protect\citeauthoryear{Bouvier et al.}{2008}]{bou} Bouvier J. et al., 2008, A\&A, 481, 661
\bibitem[\protect\citeauthoryear{Bouy et al.}{2003}]{bou03} Bouy H., Brandner W., Mart\'{i}n E.L. et al., 2003, AJ, 126, 1526
\bibitem[\protect\citeauthoryear{Brown et al.}{1989}]{bro} Brown J.A., Sneden C., Lambert D.L., Dutchover E. J., 1989, ApJS, 71, 293
\bibitem[\protect\citeauthoryear{Burgasser et al.}{2000}]{bur00} Burgasser A.J. et al., 2000, ApJ, 531, L57
\bibitem[\protect\citeauthoryear{Burgasser et al.}{2005}]{bur05} Burgasser A.J., Kirkpatrick J.D., Lowrance P.J. 2005, AJ, 129, 2849
\bibitem[\protect\citeauthoryear{Burgasser \& McElwain}{2006}]{bur06} Burgasser A.J., McElwain M.W., 2006, AJ, 131, 1007
\bibitem[\protect\citeauthoryear{Burgasser et al.}{2010}]{bur10} Burgasser A.J. et al., 2010, ApJ, submitted 
\bibitem[\protect\citeauthoryear{Burningham et al.}{2009}]{bur} Burningham B. et al., 2009, MNRAS, 395, 1237
\bibitem[\protect\citeauthoryear{Burrows et al.}{1997}]{burr97} Burrows A. et al., 1997, ApJ, 491, 856
\bibitem[\protect\citeauthoryear{Burrows et al.}{2006}]{burr06} Burrows A., Sudarsky D., Hubeny I., 2006, ApJ, 640, 1063
\bibitem[\protect\citeauthoryear{Buzzoni et al.}{1984}]{buz} Buzzoni B. et al., 1984, ESO Messenger, 38, 9
\bibitem[\protect\citeauthoryear{Caballero et al.}{2007}]{cab07} Caballero J.A., 2007, ApJ, 667, 520
\bibitem[\protect\citeauthoryear{Casali et al.}{2007}]{cas} Casali M. et al., 2007, A\&A, 467, 777
\bibitem[\protect\citeauthoryear{Casewell et al.}{2007}]{case} Casewell S.L., Dobbie P.D., Hodgkin S.T. et al., 2007, MNRAS, 378, 1131
\bibitem[\protect\citeauthoryear{Chabrier et al.}{2000}]{cha00} Chabrier G., Baraffe I., Allard F., Hauschildt P., 2000, ApJ, 542, 464
\bibitem[\protect\citeauthoryear{Chappelle et al.}{2005}]{cha} Chappelle R.J., Pinfield D.J., Steele I.A. et al., 2005, MNRAS, 361, 1323
\bibitem[\protect\citeauthoryear{Chauvin et al.}{2005}]{cha05} Chauvin G. et al., 2005, A\&A, 430, 1027
\bibitem[\protect\citeauthoryear{Chiu et al.}{2006}]{chi} Chiu K., Fan X., Leggett S.K. et al., 2006, AJ, 131, 2722
\bibitem[\protect\citeauthoryear{Clarke et al.}{2009}]{cla} Clarke J.R.A., Pinfield D.J., Burningham B. et al., 2009, arXiv: 0805.4772
\bibitem[\protect\citeauthoryear{Covey et al.}{2007}]{cov} Covey K.R. et al., 2007, AJ, 134, 2398
\bibitem[\protect\citeauthoryear{Cossburn et al.}{1997}]{cos} Cossburn M.R., Hodgkin S.T., Jameson R.F., Pinfield D.J., 1997, MNRAS, 288L, 23
\bibitem[\protect\citeauthoryear{Cruz et al.}{2003}]{cru03} Cruz K.L., Reid I.N., Liebert J. et al., 2003, AJ, 126, 2421
\bibitem[\protect\citeauthoryear{Cruz et al.}{2007}]{cru} Cruz K.L. et al., 2007, AJ, 133, 439
\bibitem[\protect\citeauthoryear{Cutri et al.}{2003}]{cut} Cutri R.M. et al., 2003, The IRSA 2MASS All-Sky Catalog of point sources (Pasadena: NASA/IPAC)
\bibitem[\protect\citeauthoryear{Day-Jones et al.}{2008}]{day} Day-Jones A.C. et al., 2008, MNRAS, 388, 838
\bibitem[\protect\citeauthoryear{da Silva et al.}{2006}]{sil} da Silva L. et al., 2006, A\&A, 458, 609
\bibitem[\protect\citeauthoryear{Deacon et al.}{2009}]{dea} Deacon N.R., Hambly N.C., King R.R., McCaughrean M.J., 2009, MNRAS, 394, 857
\bibitem[\protect\citeauthoryear{Delfosse et al.}{1997}]{del97} Delfosse X. et al., 1997, A\&A, 327, 25
\bibitem[\protect\citeauthoryear{De Simone et al.}{2004}]{sim} De Simone R., Wu X., Tremaine S., 2004, MNRAS, 350, 627
\bibitem[\protect\citeauthoryear{Ducourant et al.}{2006}]{duc} Ducourant C. et al., 2006, A\&A, 448, 1235
\bibitem[\protect\citeauthoryear{Dupuy et al.}{2009}]{dup} Dupuy T.J., Liu M.C., Ireland M.J., 2009, ApJ, 699, 168
\bibitem[\protect\citeauthoryear{Epchtein et al.}{1997}]{epc} Epchtein N., de Batz B., Capoani L. et al., 1997, ESO Messeager, 87, 27
\bibitem[\protect\citeauthoryear{Faherty et al.}{2010}]{fah10} Faherty J.K., Burgasser A.J., West A.A. et al., 2010, AJ, 139, 176
\bibitem[\protect\citeauthoryear{Famaey et al.}{2005}]{fam} Famaey B., Jorissen A., Luri X. et al., 2005, A\&A, 430, 165
\bibitem[\protect\citeauthoryear{Fan et al.}{2000}]{fan} Fan X. et al., 2000, AJ, 119, 928
\bibitem[\protect\citeauthoryear{Garc\'{i}a P\'{e}rez et al.}{2006}]{gar06} Garc\'{i}a P\'{e}rez A.E., Asplund M., Primas F. et al., 2006, A\&A, 451, 621
\bibitem[\protect\citeauthoryear{Geballe et al.}{2002}]{geb} Geballe T.R. et al., 2002, ApJ, 564, 466
\bibitem[\protect\citeauthoryear{Giclas et al.}{1971}]{gic} Giclas H.L., Burnham R., Thomas N.G., 1971, Lowell proper motion survey Northern Hemisphere. The G numbered stars. 8991 stars fainter than magnitude 8 with motions $>$ 0$^{\prime\prime}$.26/year.
\bibitem[\protect\citeauthoryear{Girardi et al.}{2000}]{gir} Girardi L., Bressan A., Bertelli G., Chiosi C., 2000, A\&AS, 141, 371
\bibitem[\protect\citeauthoryear{Gizis et al.}{2000}]{giz00} Gizis J.E., Monet D.G., Reid I.N. et al., 2000, MNRAS, 311, 385
\bibitem[\protect\citeauthoryear{Gizis et al.}{2001}]{giz01} Gizis J.E., Kirkpatrick J.D., 2001, AJ, 121, 2185
\bibitem[\protect\citeauthoryear{Golimowski et al.}{2004}]{gol04} Golimowski D.A. et al., 2004, AJ, 128, 1733
\bibitem[\protect\citeauthoryear{Gonz{\'a}lez Hern{\'a}ndez et al.}{2009}]{gon09} Gonz{\'a}lez Hern{\'a}ndez J.I., Bonifacio P., 2009, A\&A, 497, 497
\bibitem[\protect\citeauthoryear{Grevesse et al.}{2007}]{gre07} Grevesse N., Asplund M., Sauval A.J., 2007, SSR, 130, 105 
\bibitem[\protect\citeauthoryear{Gustafsson et al.}{2008}]{gus08} Gustafsson B., Edvardsson B., Eriksson K., 2008, A\&A, 486, 951 
\bibitem[\protect\citeauthoryear{Hakkila et al.}{1997}]{hak97} Hakkila J., Myers J.M., Stidham B.J. et al., 1997, AJ, 114, 2043 
\bibitem[\protect\citeauthoryear{Hambly et al.}{2001}]{ham01} Hambly N.C. et al., 2001, MNRAS, 326, 1279
\bibitem[\protect\citeauthoryear{Hambly et al.}{2008}]{ham} Hambly N.C. et al., 2008, MNRAS, 384, 637
\bibitem[\protect\citeauthoryear{Hatzes et al.}{2005}]{hat} Hatzes A.P., Guenther E.W., Endl M. et al., 2005, A\&A 437, 743
\bibitem[\protect\citeauthoryear{Hawley et al.}{2002}]{haw} Hawley S.L. et al., 2002, AJ, 123, 3409
\bibitem[\protect\citeauthoryear{Hekker \& Mel\'{e}ndez}{2007}]{hek} Hekker S., Mel\'{e}ndez J., 2007, A\&A, 475, 1003
\bibitem[\protect\citeauthoryear{Hewett et al.}{2006}]{hew} Hewett P.C., Warren S.J., Leggett S.K., Hodgkin S.T., 2006, MNRAS, 367, 454
\bibitem[\protect\citeauthoryear{Hodapp et al.}{2003}]{hod} Hodapp K.W. et al., 2003, PASP, 115, 1388
\bibitem[\protect\citeauthoryear{Hog et al.}{2000}]{hog00} Hog E. et al., 2000, A\&A, 355L, 27
\bibitem[\protect\citeauthoryear{Hogan et al.}{2008}]{hog} Hogan E., Jameson R.F., Casewell S.L. et al. 2008, MNRAS, 388, 495
\bibitem[\protect\citeauthoryear{Hurley et al.}{2000}]{hur} Hurley J.R., Pols O.R., Tout C.A., 2000, MNRAS, 315, 543
\bibitem[\protect\citeauthoryear{Ivanov}{2008}]{iva} Ivanov G.A., 2008, KFNT, 24, 480
\bibitem[\protect\citeauthoryear{Jenkins et al.}{2009}]{jen} Jenkins J. et al., 2009, MNRAS, 398, 911
\bibitem[\protect\citeauthoryear{Johnson et al.}{2009}]{joh09} Johnson J.A., Apps K., 2009, ApJ, 699, 933
\bibitem[\protect\citeauthoryear{Kendall et al.}{2007}]{ken07} Kendall T.R., Jones H.R.A., Pinfield D.J. et al., 2007, MNRAS, 374, 445
\bibitem[\protect\citeauthoryear{Kharchenko}{2001}]{kh1} Kharchenko N.V., 2001, KFNT,  17, 409
\bibitem[\protect\citeauthoryear{Kharchenko et al.}{2004}]{kha} Kharchenko N.V., Piskunov A.E., R\"{o}ser S. et al., 2004, AN, 325, 740
\bibitem[\protect\citeauthoryear{Kirkpatrick et al.}{1999}]{kir99} Kirkpatrick J.D. et al., 1999, ApJ, 519, 802
\bibitem[\protect\citeauthoryear{Kirkpatrick et al.}{2000}]{kir00} Kirkpatrick J.D. et al., 2000, AJ, 120, 447
\bibitem[\protect\citeauthoryear{Knapp et al.}{2004}]{kna04} Knapp G.R. et al., 2004, AJ, 127, 3553
\bibitem[\protect\citeauthoryear{Kroupa}{1995}]{kro95} Kroupa P., 1995, ApJ, 453, 350
\bibitem[\protect\citeauthoryear{Lawrence et al.}{2007}]{law} Lawrence A. et al., 2007, MNRAS, 379, 1599
\bibitem[\protect\citeauthoryear{L\'{e}pine \& Shara}{2005}]{lep} L\'{e}pine S., Shara M.M., 2005, AJ, 129, 1483
\bibitem[\protect\citeauthoryear{L\'{e}pine}{2005}]{lea} L\'{e}pine S., 2005, AJ, 130, 1680
\bibitem[\protect\citeauthoryear{Liu et al.}{2006}]{lium} Liu M.C., Leggett S.K., Golimowski D.A. et al., 2006, ApJ, 647, 1393
\bibitem[\protect\citeauthoryear{Liu et al.}{2008a}]{liu8} Liu M.C., Dupuy T.J., Ireland M.J., 2008a, ApJ, 689, 436
\bibitem[\protect\citeauthoryear{Liu et al.}{2008b}]{liu} Liu Y.-J. et al., 2008b, ApJ, 672, 553
\bibitem[\protect\citeauthoryear{Lovis \& Mayor}{2007}]{lov} Lovis C., Mayor M., 2007, A\&A, 472, 657
\bibitem[\protect\citeauthoryear{Lowrance et al.}{1999}]{low99} Lowrance P.J. et al., 1999, ApJ, 512, L69
\bibitem[\protect\citeauthoryear{Lowrance et al.}{2000}]{low00} Lowrance P.J. et al., 2000, ApJ, 541, 390
\bibitem[\protect\citeauthoryear{Luck \& Heiter}{2007}]{luc} Luck R.E., Heiter U., 2007, AJ, 133, 2464
\bibitem[\protect\citeauthoryear{Luhman et al.}{2007}]{luh} Luhman K.L. et al., 2007, ApJ, 654, 570
\bibitem[\protect\citeauthoryear{Luyten \& Hughes}{1980}]{luy} Luyten W.J., Hughes H.S., 1980, Proper motion survey with the forty-eight inch Schmidt telescope. LV. First supplement to the NLTT catalogue (Mineapolis, NM: Univ. of Minnesota).
\bibitem[\protect\citeauthoryear{Magazz\`{u} et al.}{1993}]{mag93} Magazz\`{u} A., Mart\'{i}n E.L., Rebolo R., 1993, ApJ, 404, L17
\bibitem[\protect\citeauthoryear{Magazz\`{u} et al.}{1998}]{mag} Magazz\`{u} A., Rebolo R., Zapatero Osorio M.R. et al., 1998, ApJ, 497, L47
\bibitem[\protect\citeauthoryear{Mart\'{i}n et al.}{1996}]{mar96} Mart\'{i}n E.L., Rebolo R., Zapatero Osorio, M.R.,1996, ApJ, 469, 706
\bibitem[\protect\citeauthoryear{Mart\'{i}n et al.}{1998}]{mar98} Mart\'{i}n E.L., Basri G., Zapatero Osorio M.R., Rebolo R., Garc\'{i}a L\`{o}pez R.J.,1998, ApJ, 507, L41
\bibitem[\protect\citeauthoryear{Metchev et al.}{2006}]{met06} Metchev S.A., Hillenbrand L.A. 2006, ApJ, 651, 1166
\bibitem[\protect\citeauthoryear{Monet et al.}{2003}]{mon} Monet D.G. et al., 2003, AJ, 125, 984
\bibitem[\protect\citeauthoryear{Montagnier et al.}{2006}]{mon06} Montagnier G. et al., 2006, A\&A, 460, L19
\bibitem[\protect\citeauthoryear{Mugrauer et al.}{2007}]{mug07} Mugrauer M., Seifahrt A., Neuh\"{a}user R. 2007, MNRAS, 378, 1328
\bibitem[\protect\citeauthoryear{Neuh\"{a}user et al.}{2004}]{neu04} Neuh\"{a}user R., Guenther E.W., 2004, A\&A, 420, 647
\bibitem[\protect\citeauthoryear{Neuh\"{a}user et al.}{2005}]{neu05} Neuh\"{a}user R., Guenther E.W., Wuchterl G. et al., 2005, A\&A, 435, L13
\bibitem[\protect\citeauthoryear{Niedzielski et al.}{2009}]{nie} Niedzielski A., Nowak G., Adam\'{o}w M., Wolszczan A. 2009, arXiv: 0906.1804
\bibitem[\protect\citeauthoryear{Phan-Bao et al.}{2008}]{pha08} Phan-Bao N. et al., 2008, MNRAS, 383, 831
\bibitem[\protect\citeauthoryear{Perryman et al.}{1997}]{per} Perryman M.A.C. et al., 1997, A\&A, 323, L49
\bibitem[\protect\citeauthoryear{Pinfield et al.}{2006}]{pin} Pinfield D.J., Jones H.R.A., Lucas P.W. et al., 2006, MNRAS, 368, 1281
\bibitem[\protect\citeauthoryear{Pinfield et al.}{2008}]{pinf} Pinfield D.J., Burningham B., Tamura M. et al., 2008, MNRAS, 390, 304
\bibitem[\protect\citeauthoryear{Radigan et al.}{2008}]{rad08} Radigan J. Lafreni\`{e}re D., Jayawardhana R., Doyon R.  2008, ApJ, 689, 471 
\bibitem[\protect\citeauthoryear{Radigan et al.}{2009}]{rad09} Radigan J., Lafreni\`{e}re D., Jayawardhana R., Doyon R. 2009, ApJ, 698, 405
\bibitem[\protect\citeauthoryear{Ram{\'i}rez et al.}{2005}]{ram05} Ram{\'i}rez I., Mel\'{e}ndez J.,  2005, ApJ, 626, 465
\bibitem[\protect\citeauthoryear{Rebolo et al.}{1995}]{reb} Rebolo R., Zapatero Osorio M.R.Z., Mart\'{i}n E.L., 1995, Nature, 377, 129
\bibitem[\protect\citeauthoryear{Rebolo et al.}{1998}]{reb98} Rebolo R., Zapatero Osorio M.R., Madruga S. et al., 1998, Science, 282, 1309
\bibitem[\protect\citeauthoryear{Reid et al.}{2006}]{rei06} Reid I.N., Walkowicz L.M., 2006, PASP, 118, 671
\bibitem[\protect\citeauthoryear{Reid et al.}{2008}]{rei08} Reid I.N., Cruz K.L., Kirkpatrick J.D. et al., 2008, AJ, 136, 1290
\bibitem[\protect\citeauthoryear{Ribas}{2003}]{rib} Ribas I., 2003, A\&A, 400, 297
\bibitem[\protect\citeauthoryear{R\"{o}ser et al.}{2008}]{ros} R\"{o}ser S., Schilbach E., Schwan H. et al., 2008, A\&A, 488, 401
\bibitem[\protect\citeauthoryear{Ruiz et al.}{1997}]{rui97} Ruiz M.T., Leggett S.K., Allard F., 1997, ApJ, 491, L107
\bibitem[\protect\citeauthoryear{Salim \& Gould}{2003}]{sal} Salim S., Gould A. 2003, ApJ, 582, 1011
\bibitem[\protect\citeauthoryear{Schneider et al.}{2002}]{sch02} Schneider D.P., et al. 2002, AJ, 123, 458
\bibitem[\protect\citeauthoryear{Scholz et al.}{2003}]{sch03} Scholz R.D., McCaughrean M.J., Lodieu N., Kuhlbrodt B., 2003, A\&A, 398, L29
\bibitem[\protect\citeauthoryear{Scholz et al.}{2004}]{sch04} Scholz Lodieu N., McCaughrean M.J. 2004, A\&A, 428, L25
\bibitem[\protect\citeauthoryear{Seifahrt et al.}{2005a}]{sei05a} Seifahrt A., Guenther E., Neuh\"{a}user, 2005a, A\&A, 440, 967
\bibitem[\protect\citeauthoryear{Seifahrt et al.}{2005b}]{sei05b} Seifahrt A., Mugrauer M., Wiese M. et al., 2005b, AN, 326, 974
\bibitem[\protect\citeauthoryear{Sheppard \& Cushing}{2009}]{she} Sheppard S.S., Cushing M.C., 2009, AJ, 137, 304
\bibitem[\protect\citeauthoryear{Skrutskie et al.}{2006}]{skr} Skrutskie  M.F. et al., 2006, AJ, 131, 1163
\bibitem[\protect\citeauthoryear{Stoughton et al.}{2002}]{sto} Stoughton C. et al., 2002, AJ, 123, 485
\bibitem[\protect\citeauthoryear{Taylor}{2006}]{tay06} Taylor B.J. 2006, AJ, 132, 2453
\bibitem[\protect\citeauthoryear{Tinney et al.}{1997}]{tin97} Tinney C.G., Delfosse X., Forveille T., 1997, ApJ, 490, L95
\bibitem[\protect\citeauthoryear{Tinney}{1998}]{tin} Tinney C.G., 1998, MNRAS, 296, L42
\bibitem[\protect\citeauthoryear{Tokovinin et al.}{2006}]{tok} Tokovinin A., Thomas S., Sterzik M., Udry S., 2006, A\&A, 450, 681
\bibitem[\protect\citeauthoryear{van Leeuwen}{2007}]{van} van Leeuwen F., 2007, A\&A, 474, 653
\bibitem[\protect\citeauthoryear{van Leeuwen}{2009}]{va2} van Leeuwen F., 2009, A\&A, 497, 209
\bibitem[\protect\citeauthoryear{West et al.}{2008}]{wes} West A.A., Hawley S.L., Bochanski J. J., 2008, AJ, 135, 785
\bibitem[\protect\citeauthoryear{White et al.}{1999}]{whi99} White R.J., Ghez A.M., Reid I.N., Schultz G., 1999, ApJ, 520, 811
\bibitem[\protect\citeauthoryear{Wilson et al.}{2001}]{wil01} Wilson J.C., Kirkpatrick J.D., Gizis J.E. et al., 2001, AJ, 122, 1989
\bibitem[\protect\citeauthoryear{Yi et al.}{2001}]{yi} Yi S., Demarque P., Kim Y.-C. et al., 2001, ApJS, 136, 417
\bibitem[\protect\citeauthoryear{York et al.}{2000}]{yor} York D.G. et al., 2000, AJ, 120, 1579
\bibitem[\protect\citeauthoryear{Zacharias et al.}{2004}]{zac04} Zacharias N., Monet D.G., Levine S.E. et al., 2004, AAS, 205, 4815
\bibitem[\protect\citeauthoryear{Zacharias et al.}{2009}]{zac09} Zacharias N. et al., 2009, yCat, 1315, 0
\bibitem[\protect\citeauthoryear{Zapatero Osorio et al.}{1998}]{zap} Zapatero Osorio M.R. et al., 1998, Brown dwarfs and extrasolar planets, P51
\bibitem[\protect\citeauthoryear{Zhang et al.}{2009a}]{zh09a} Zhang Z.H., Jones H.R.A., Pinfield D.J. et al., 2009a, Proceedings of the 10th Asian-Pacific Regional IAU Meeting 2008, P191, arXive: 0902.2677
\bibitem[\protect\citeauthoryear{Zhang et al.}{2009b}]{zh09b} Zhang Z.H. et al., 2009b, A\&A, 497, 619

\end{thebibliography}
\end{document}